\documentclass[hidelinks]{article}

\usepackage{arxiv}

\usepackage[utf8]{inputenc} % allow utf-8 input
\usepackage[T1]{fontenc}    % use 8-bit T1 fonts
\usepackage{amsmath}
\usepackage{amssymb}
\usepackage{graphicx}
\usepackage{lipsum}
\usepackage{siunitx}
\usepackage{physics}
\usepackage{xspace}
\usepackage{multirow}
\usepackage{hyperref}
\usepackage{lineno}
\usepackage[round,authoryear]{natbib}
\usepackage[nameinlink, capitalize]{cleveref}

\author{
 Ethan YoungIn Shin$^1$ and Michael F. Howland$^{1*}$ \\
 \\
  $^{1}$Civil and Environmental Engineering, Massachusetts Institute of Technology, Cambridge, MA 02139, USA \\
  $^*$Corresponding author: \texttt{mhowland@mit.edu} \\
}

\title{Probabilistic inference of surface parameters for Monin--Obukhov similarity theory}

\begin{document}

\maketitle

\begin{abstract}
In simulations of atmospheric flow, the grid spacing typically exceeds the size of the roughness elements at the surface by an order of magnitude. 
The unresolved effects of surface morphology and roughness on the flow are represented by effective surface parameters and specified as a surface flux boundary condition, most often through a formulation based on Monin--Obukhov similarity theory (MOST). 
These surface parameters are known to depend on both surface and flow properties, yet they are generally estimated as deterministic quantities, either by a roughness model or by measurement-based anemometric approaches, with no characterization of associated uncertainty. 
In this study, we use a Bayesian approach to infer surface parameters for MOST and quantify the uncertainty associated with them. 
For the aerodynamic roughness length $z_0$ inferred in isolation, a normal--normal conjugate update yields the posterior and posterior predictive distributions in closed form. 
We first demonstrate the probabilistic $z_0$ estimation method on idealized conventionally neutral boundary layers generated by large-eddy simulation, where $z_0$ is prescribed, and quantify how prior- and observation-related choices shape the inferred posterior. 
We then apply the method to field observations from the Atmospheric Radiation Measurement Southern Great Plains observatory, from which we infer site-specific posterior distributions of $z_0$ conditioned on month, on wind direction, and on both jointly. 
By leveraging a prior informed by sample statistics of all near-neutral observations in the training years, we demonstrate the advantage of the Bayesian inference method for $z_0$ relative to the state-of-practice least-squares profile-fitting method in conditions of data sparsity.
Predictions for unseen observations in a test year---evaluated at the posterior mean---reduce root mean squared error and mean absolute error in data-sparse wind directions, while the posterior predictive distributions consistently reduce the continuous ranked probability score by approximately $20$--$30\%$ across all wind directions considered.
\end{abstract}

\section*{Keywords}

Surface parameters, Bayesian inference, Large-eddy simulation, Field observations, Surface--atmosphere interactions

\maketitle
% \tableofcontents

\section{Introduction}
\label{sec:introduction}

% WHY SURFACE PARAMETERS ARE WORTH STUDYING
The atmospheric boundary layer (ABL) is the region of the troposphere that directly responds to surface forcings on timescales of an hour or less \citep{stull2012introduction}.
The combined effects of surface heterogeneity and the wide range of dynamically relevant flow scales---from large mesoscale motions down to the viscous length scale---mean that the surface layer of the ABL is unlikely to ever be fully resolved in atmospheric models \citep{mahrt2000surface,bou2020persistent}.
Instead, these effects are parameterized by surface parameterizations, which rely on assumptions of local flow homogeneity and statistical stationarity, upon which classical scaling laws for the ABL are built \citep{foken200650}.
In the surface layer, mean wind speed varies approximately logarithmically with height under neutral conditions, and Monin--Obukhov similarity theory (MOST) extends this framework to thermally stratified regimes \citep{monin1954basic}.
Within this framework, the aerodynamic roughness length $z_0$---defined as the extrapolated height at which the logarithmic wind profile reaches zero---represents the integrated effect of unresolved surface roughness elements on the flow aloft \citep{wiernga1993representative, grimmond1999aerodynamic}.
This physical parameter spans several orders of magnitude across geophysical land cover types \citep{stull2012introduction}.
In canopy flows, the zero-plane displacement distance $d$ shifts the origin of the logarithmic profile upward and scales primarily with canopy height, reaching values of $d \sim \mathcal{O}(10~\mathrm{m})$ for tall vegetation and urban canopies \citep{jackson1981displacement, anderson2010large, sogachev2016displacement}.
Surface parameters are key inputs to operational weather and climate models, and their accurate estimation is essential for reliable modeling and prediction of winds in the ABL \citep{andre1986effective,bou2007parameterization}.

% LITERATURE REVIEW AND GENERAL STATE OF RESEARCH
Estimating surface parameters for geophysical flows has been an active area of research for several decades.
Early methods relied on the geometric properties of roughness elements \citep{lettau1969note}.
Profile-based (anemometric) least-squares approaches were developed for wind speed measurements at multiple levels \citep{bergeron1992estimating} and at a single level \citep{martano2000estimation}.
Gustiness-based methods, which require only standard wind speed and gust records, offered a practical alternative for sites where multi-level wind profiles are unavailable \citep{verkaik2000evaluation}.
More recently, increased data diversity and availability have motivated estimation methods based on satellite measurements \citep{prigent2005estimation, floors2021satellite}, offshore wind measurements \citep{golbazi2019methods}, and LiDAR profiling measurements \citep{hammond2012roughness}. 
Eddy covariance flux tower networks have also provided a basis for site-specific estimation of both $z_0$ and $d$ \citep{horne2026estimating}, though the robustness of $d$ estimates remains a known challenge \citep{graf2014intercomparison}.
For the anemometric approaches, least-squares fitting of the log law to wind speed observations remains the state of practice, yet it returns a deterministic estimate with no accompanying measure of uncertainty, despite the known sensitivity of $z_0$ to choices made in preparing the observations \citep{barthelmie1993estimation, he2017estimation}.

% RESEARCH GAP
Recovering surface parameters from wind speed observations is an inverse problem, and one that may be poorly conditioned.
Because wind speed depends logarithmically on $z_0$, small perturbations in observations translate into large changes in inferred roughness.
Turbulence introduces variability into wind measurements that finite averaging cannot eliminate \citep{wyngaard2010turbulence}.
Statistical convergence typically requires extending the averaging window until the residual fluctuation is negligible.
However in the ABL, boundary conditions and forcing evolve on timescales comparable to those required for convergence, so extending the window eventually admits non-stationarity faster than it suppresses turbulent variability \citep{metzger2008time,mahrt2020non}.
Every observation is therefore a finite-time average carrying an unavoidable sample error.
This sensitivity is compounded by heuristic choices for the observations used in the fitting process, including the number of observations and the range of heights used in the fit, both of which significantly affect the result \citep{barthelmie1993estimation,he2017estimation}.
A deeper difficulty is that, despite the name, surface parameters are not properties of the surface alone.
Previous works have reported $z_0$ to also depend on flow conditions such as atmospheric stability \citep{bou2007parameterization, zilitinkevich2008effect, bou2020persistent}.
Fundamentally, surface parameters represent unresolved surface--atmosphere interactions that are spatiotemporally variable, so a single fixed value cannot represent them faithfully---yet nearly all existing estimation methods yield deterministic point estimates that do not formally quantify the associated uncertainties.
Here a Bayesian approach is better matched to the problem: rather than seeking a fixed-point estimate, it treats $z_0$ as an uncertain quantity whose distribution reflects the uncertainty in the observations and can be updated as new observations become available.
Because roughness lengths cannot be measured directly and reported estimates are known to exhibit discrepancies by about a factor of two \citep{wiernga1993representative}, quantifying and communicating this uncertainty creates opportunities for error reduction and better-informed decision-making in operational models.

% THIS STUDY'S OBJECTIVES
This study introduces a Bayesian approach to infer surface parameters for MOST, which remains the basis for surface-layer parameterizations in operational weather and climate models.
The method yields closed-form expressions for the posterior and posterior predictive distributions, which quantify the uncertainty in the inferred $z_0$ and the resulting uncertainty in predictions of unseen observations, respectively.
The method is applied to two settings: synthetic observations from idealized ABL flows generated by large-eddy simulation to verify the methodology, and observations from a field measurement campaign site.
In the idealized ABL flow setting, we provide proof-of-concept and quantify how the inferred uncertainty in $z_0$ depends on prior- and observation-related choices.
In the field setting, we apply the method to infer site-specific probabilistic estimates of $z_0$ from wind speed observations and give recommendations for its use in practice.

% METHODOLOGY USED IN THIS STUDY
\section{Methodology}
\label{sec:methodology}

This section presents the wind profile model based on MOST and formulates the Bayesian inverse problem for inferring surface parameters from normalized wind speed observations using the model.
In this study, we provide two proof-of-concept tests where we take observations from two sources: idealized ABL flows generated by large-eddy simulation for verification, and field measurements collected from an eddy covariance flux system at a field measurement campaign site.

\subsection{Wind profile model based on Monin--Obukhov similarity theory}
\label{sec:windprofilemodel}

In MOST, the wind speed profile is formulated as 
\begin{equation}
    \langle U \rangle =\frac{\langle u_* \rangle}{\kappa} \left[ \log \left( \frac{z-d}{z_0} \right) - \psi_m \left( \frac{z}{\langle L_o \rangle} \right) \right]
    \label{eq:loglaw_most}
\end{equation}
where the angled brackets $\langle \cdot \rangle$ denote ensemble averaging, $U$ is the horizontal wind speed, $u_*$ is the friction velocity, $\kappa$ is the von K\'{a}rm\'{a}n constant, $z$ is the vertical height, $d$ is the displacement distance, $z_0$ is the roughness length, $\psi_m$ is the stability correction function for momentum, and $L_o$ is the Obukhov length.
The similarity theory is built on dimensional analysis and key assumptions that include fully developed turbulence, statistical quasi-stationarity of the flow, and horizontally homogeneous surface conditions \citep{wyngaard2010turbulence}.

The effects of atmospheric stability are incorporated through the stability correction function $\psi_m$, which uses the Obukhov length $L_o$.
The mean Obukhov length is defined as
\begin{equation}
    \langle L_o \rangle  = -\,\frac{ \langle u_*\rangle^3 \, \langle \theta_s \rangle }{\kappa g \, (\langle w'\theta' \rangle_s)},
    \label{eq:obukhovlength}
\end{equation}
where $\langle \theta_s \rangle$ is the mean surface potential temperature, $g$ is the gravitational acceleration, and $\langle w'\theta' \rangle_s$ denotes the kinematic surface heat flux.

We prescribe the stability correction function for momentum $\psi_m$ as
\begin{equation}
  \psi_m( z / \langle L_o \rangle) =
  \begin{cases}
    -\beta z / \langle L_o \rangle, & z / \langle L_o \rangle \geq 0, \\[6pt]
    2\log\!\left(\dfrac{1+x}{2}\right)
    + \log\!\left(\dfrac{1+x^2}{2}\right)
    - 2\arctan x + \dfrac{\pi}{2}, & z / \langle L_o \rangle < 0,
  \end{cases}
  \label{eq:stabilityfunction}
\end{equation}
with $x = (1 - \gamma z / \langle L_o \rangle)^{1/4}$. 
The coefficients $\beta=4.7$ and $\gamma=15$ are the empirical constants of the non-integrated Businger-Dyer functions $\phi_m = 1 + \beta z / \langle L_o \rangle$ and $\phi_m = (1 - \gamma z / \langle L_o \rangle)^{-1/4}$, respectively \citep{businger1971flux,dyer1974review}.

Together, Eqs.~\ref{eq:loglaw_most}, \ref{eq:obukhovlength} and \ref{eq:stabilityfunction} constitute the MOST-based wind profile model, which maps the inputs (vertical height, friction velocity, surface parameters, atmospheric stability) to the output (wind speed).
In the next section, we describe how applying Bayes’ theorem enables us to combine a prior with this model to conduct probabilistic inference of surface parameters.

\subsection{Probabilistic inference of surface parameters}
\label{sec:probabilisticinference}

We formulate the inference of surface parameters from wind speed observations as a Bayesian inverse problem, with the MOST-based wind profile as the forward model.
First we demonstrate that inferring the aerodynamic roughness length $z_0$ in isolation, taking the displacement distance $d$ and the tuned coefficients in $\psi_m$ as known, results in closed-form expressions for the posterior and posterior predictive distributions.
In Appendix~\ref{sec:appendix2}, we describe a numerical method for estimating the joint probability distribution over multiple unknown parameters in MOST from wind speed observations at several vertical levels.

Here we take observations of wind speed and friction velocity at a single vertical level $z$ over $n$ independent instantiations.
Each instantiation is an observation averaged over a non-overlapping time window of length $\Delta t_{\text{avg}}$.
Invoking ergodicity, the time average approximates the ensemble average.
A complete observation set is therefore the vector $y \in \mathbb{R}^{n}$ of time-averaged, friction-velocity-normalized wind speeds, with entries
\begin{equation}
    y_{i} = \frac{\langle U \rangle_{i}}{\langle u_* \rangle_i}, \qquad i=1,\dots,n.
    \label{eq:observation}
\end{equation}
For inferring $z_0$, we retain only neutral or near-neutral observations, so that the inferred values do not depend on the empirical stability functions.
Normalizing by $\langle u_* \rangle_i$ collapses the observations onto the logarithmic profile, and the forward model depends only on the measurement height $z$ and the roughness length
\begin{equation}
    f(z_0) = \frac{1}{\kappa}\log\!\left(\frac{z}{z_0}\right).
    \label{eq:forwardmodel}
\end{equation}

Bayes' theorem relates the posterior distribution of $z_0$ to the likelihood and prior,
\begin{equation}
    p(z_0 \mid y) \propto p(y \mid z_0)\, p(z_0),
    \label{eq:bayes}
\end{equation}
where the likelihood $p(y \mid z_0)$ is the probability of the observations under the forward model.
We model the $n$ observations $y=[y_1,...,y_n]^\top$ as conditionally independent given $z_0$, each sharing the same observational noise variance $\sigma^2$, which we estimate from the sample variance of the observations and hold fixed throughout the update.
The joint likelihood is then
\begin{equation}
    p(y \mid z_0) = \prod_{i=1}^n \mathcal{N}\!\left(y_i \mid f(z_0),\, \sigma^2\right).
    \label{eq:likelihood}
\end{equation}

In choosing a prior, we consider the physical constraint that $z_0$ must be positive.
We enforce this by introducing a variable $\phi$ from a transformed, unconstrained, normally distributed space $\mathcal{N}(\mu_0, \tau_0^2)$, where $\mu_0$ and $\tau_0^2$ are the prior mean and variance, respectively.
This distribution can be transformed back to the physical, constrained space,
\begin{equation}
    \phi = \log(z_0) \iff z_0 = e^{\phi},
    \label{eq:logtransformation}
\end{equation}
where $z_0$ is now strictly positive by construct of a lognormal prior distribution
\begin{equation}
    p(z_0) = \mathrm{lognormal}(z_0 \mid \mu_0,\,\tau_0^2).
\end{equation}
The transformation also renders the forward model linear in $\phi$,
\begin{equation}
    y_i = \frac{1}{\kappa}\log z - \frac{1}{\kappa}\,\phi + \varepsilon_i,
    \qquad \varepsilon_i \sim \mathcal{N}(0,\sigma^2).
    \label{eq:linearmodel}
\end{equation}
The transformed prior $p(\phi) = \mathcal{N}(\phi \mid \mu_0,\tau_0^2)$ and the transformed likelihood are both normally distributed, and the normal--normal conjugate update yields a closed-form posterior distribution.
Full details of the derivation may be found in \citet{gelman1995bayesian}.
The sample mean $\overline{y} = \frac{1}{n}\sum_i^n y_i$ is a sufficient statistic for $\phi$, and the conjugate update after $n$ observations gives the posterior mean and variance
\begin{equation}
    \mu_n = \tau_n^2\!\left(\frac{\mu_0}{\tau_0^2}
        - \frac{n}{\kappa \sigma^2} \left( \overline{y} - \frac{1}{\kappa} \log z \right) \right), \quad
    \tau_n^2 = \left(\frac{1}{\tau_0^2}
        + \frac{n}{\kappa^2 \sigma^2}\right)^{-1}.
    \label{eq:posterior_mean_variance}
\end{equation}
Mapping back through Eq.~\ref{eq:logtransformation} recovers the closed-form posterior for $z_0$,
\begin{equation}
    p(\phi \mid y) = \mathcal{N}(\phi \mid \mu_n,\tau_n^2)
    \iff
    p(z_0 \mid y) = \mathrm{lognormal}(z_0 \mid \mu_n,\tau_n^2).
    \label{eq:posteriordistribution}
\end{equation}

Here we make explicit the relationship between the posterior (Eq.~\ref{eq:posteriordistribution}) and the maximum likelihood estimate (MLE) to draw theoretical insight into the comparison between our method and least-squares regression, which coincides exactly with MLE when errors are assumed to be independent and identically distributed normal variables.
Maximizing the likelihood (Eq.~\ref{eq:likelihood}) with respect to $\phi$ gives $\hat\phi = \log z - \kappa \overline{y}$, whose sampling distribution over $n$ observations is
\begin{equation}
    \hat\phi \sim \mathcal{N}\!\left( \phi,\ \frac{\kappa^2\sigma^2}{n} \right).
    \label{eq:mlesampling}
\end{equation}
Introducing the terms, prior and data precisions $\lambda_0 = 1/\tau_0^2$ and $\lambda_n = n/\kappa^2\sigma^2$, the conjugate update (Eq.~\ref{eq:posterior_mean_variance}) may be written as the precision-weighted average
\begin{equation}
    \mu_n = \frac{\lambda_0 \mu_0 + \lambda_n \hat\phi}{\lambda_0 + \lambda_n},
    \qquad
    \tau_n^2 = \left( \lambda_0 + \lambda_n \right)^{-1},
    \label{eq:precisionform}
\end{equation}
which shows the posterior mean as an interpolation between prior mean and MLE, and the posterior precision as the sum of prior and data precisions.
It also shows that the posterior and the sampling distribution of the MLE become indistinguishable in two limits.
As $\tau_0 \rightarrow \infty$ with $n$ fixed, the posterior (Eq.~\ref{eq:posteriordistribution}) becomes the MLE sampling distribution (Eq.~\ref{eq:mlesampling}) exactly, since $\mu_n \rightarrow \hat\phi$ and $\tau_n^2 \rightarrow \kappa^2\sigma^2/n$.
As $n \rightarrow \infty$ with $\tau_0$ fixed, they agree asymptotically, where the posterior mean differs with MLE by $\mu_n - \hat\phi = \mathcal{O}\!\left( 1/n \right)$, and the posterior variance shrinks as $\mathcal{O}(1/n)$.
This informs us that the posterior variance of $\phi=\log z_0$, the learned uncertainty in transformed space, will decrease monotonically with $n$ and approach $\kappa^2\sigma^2/n$ as the estimate becomes increasingly determined by the data rather than the prior.
Conversely, the prior will remain important when the data are sparse (small $n$) or noisy (large $\sigma^2$)---a condition typical of measurement-based applications.
We verify this behavior in synthetic observations and then leverage it in field observations to improve out-of-sample predictions.

To predict unseen observations $\tilde{y}$, we require the posterior predictive distribution, $p(\tilde y \mid y)$.
Under the normal--normal conjugate update, the posterior for the unseen $\tilde\phi = \log \tilde z_0$ is available analytically as
\begin{equation}
    p(\tilde\phi \mid y) = \mathcal{N}\!\left( \tilde \phi \mid \mu_n,\ \tau_n^2\right).
\end{equation}
The observation model is affine in $\tilde\phi$, with conditional mean $\operatorname{E}(\tilde y \mid \tilde\phi) = (\log z - \tilde\phi)/\kappa$ and independent observational error of variance $\sigma^2$, so that $\tilde y = (\log z - \tilde\phi)/\kappa + \eta$ with $\eta \sim \mathcal{N}(0, \sigma^2)$.
This affine mapping shifts the posterior mean to $(\log z - \mu_n)/\kappa$ and scales the posterior variance by $1/\kappa^2$, and the observational error adds a further $\sigma^2$ to the variance, which gives the closed-form posterior predictive distribution
\begin{equation}
    p(\tilde y \mid y) = \mathcal{N}\!\left(\tilde y \,\middle|\,
    \frac{\log z - \mu_n}{\kappa},\ \frac{\tau_n^2}{\kappa^2} + \sigma^2\right).
    \label{eq:posteriorpredictivedistribution}
\end{equation}
Because the forward model is affine in $\phi$, the precisions $\lambda_0$ and $\lambda_n$ carry into the predictions.
The posterior predictive mean is the same weighted average of the least-squares estimate $\hat \phi$ and the prior mean $\mu_0$ through $\mu_n$ (see Eq.~\ref{eq:precisionform}).
This means that with more observations or low observational noise, the posterior predictive mean and the least-squares prediction coincide.
Conversely, when $n$ is small or $\sigma^2$ is large, the prior will cause a difference of $(\mu_n - \hat\phi)/\kappa$ between the two predictions of the mean.
Similarly, with more observations or low observational noise, the posterior predictive width converges to the observation noise, while at small $n$ or large $\sigma^2$, it widens to reflect the additional uncertainty in the inferred $z_0$.

To summarize, our Bayesian method quantifies uncertainty in both the aerodynamic roughness length $z_0$ (Eq.~\ref{eq:posteriordistribution}) and predictions of unseen wind speed observations (Eq.~\ref{eq:posteriorpredictivedistribution}), conditioned on the observed data.
Both quantities are obtained at negligible computational cost relative to the deterministic least-squares fits in current use.

\subsection{Observations from large-eddy simulations}
\label{sec:observations_les}

For large-eddy simulation (LES) of high-Reynolds-number atmospheric flows, the near-wall region (including the viscous and buffer layers, and in many cases part of the logarithmic layer) is not explicitly resolved and is instead represented using a so-called rough wall model, which uses the aerodynamic roughness length $z_0$ \citep{moeng1984large,kumar2006large,stoll2020large}.
This wall model for LES of atmospheric flows is commonly formulated using MOST \citep{moeng1984large, bou2005scale}.
Because $z_0$ is prescribed directly in the LES wall model, its true value is known.
This provides a controlled setting for proof-of-concept of the Bayesian inversion method.
The known ground truth further enables a systematic quantification of how uncertainty in the inferred $z_0$ depends on prior- and observation-related choices.

In this study, the incompressible flow solver Pad\'{e}Ops\footnote{\url{https://github.com/Howland-Lab/PadeOps}} \citep{ghate2017subfilter, howland2020influence} is used to solve the filtered, incompressible Navier–Stokes equations on a uniformly discretized, staggered grid.
Further details of the solver may be found in \citet{heck2025coriolis}.
Here we use the sigma subgrid-scale model \citep{nicoud2011using} and the wall model developed by \citet{bou2005scale}, which provides a local shear-stress boundary condition computed from MOST.
We choose matching velocities, the input velocities used by the wall model to predict the surface shear stress, to be evaluated at the first staggered grid point above the surface.
To investigate the inference of the aerodynamic roughness length in the neutral atmospheric surface layer, we simulate flow in the conventionally neutral boundary layer, using a setup by \citet{liu2021geostrophic}.
The use of the solver for this conventionally neutral boundary layer setup has been validated and used in previous studies \citep{heck2025coriolis, shin2025addressing}.
The computational domain size is set to $[2\pi \times 2\pi \times 2]~\mathrm{km}$ in the streamwise ($x$), spanwise ($y$), and vertical ($z$) directions, respectively.
The domain is discretized using $72$ grid points in each spatial direction.
Periodic boundary conditions are used in streamwise and spanwise directions, while a Neumann boundary condition for the surface shear stress is applied locally by the wall model.
The imposed geostrophic wind speed is $12~\mathrm{m~s^{-1}}$ and the initial potential temperature profile is $\theta(z) = \theta_0 + \Gamma z$ where $\theta_0=300~\mathrm{K}$ and the free-atmospheric lapse rate is $\Gamma=3~\mathrm{K~km^{-1}}$.
The latitude is $\phi=70^\circ$, which corresponds to the Coriolis parameter $f = 2 \Omega \sin \phi = 1.37 \times 10^{-4}$ where $\Omega=7.29\times 10^{-5}~\mathrm{rad~s^{-1}}$ is the Earth's angular rotation rate.
We spin the simulation up for approximately $9 / f \approx 20~\mathrm{h}$ and then take our observations to be time-averaged statistics, sampled at a single $xyz$-location within one inertial period $2\pi/f \approx 13~\mathrm{h}$ to minimize the effects of inertial oscillations \citep{liu2021geostrophic}.

\begin{table}
\centering
\caption{Parametric study of the effects of the prior mean of $z_0$ and the number and time-averaging length of observations on the inferred $z_0$ posterior}
\renewcommand{\arraystretch}{1.2}
\begin{tabular}{lc}
\hline
Factor & Values \\ 
\hline
Prior mean of $z_0$ [m] & $0.01$, $1$ \\
Number of observations $n$ [-] & $1$, $3$, $5$, $7$, $9$ \\
Time-averaging length $\Delta t_{\text{avg}}$ [s] & 5, 10, 30, 60, 300, 600, 1800, 3600 \\
\hline
\end{tabular}
\label{tab:parametricstudy}
\end{table}

The idealized testbed serves two purposes: to demonstrate proof-of-concept of the method, and to quantify how prior- and observation-related choices affect the inferred roughness length.
Here we consider the prescribed roughness length $z_0=0.01~\mathrm{m}$, which serves as the ground truth and is representative of fairly level grass plains \citep{stull2012introduction}.
For the prior-related choice, we consider two prior means of $z_0$, one at the truth ($0.01~\mathrm{m}$) and one displaced by two orders of magnitude ($1~\mathrm{m}$).
For observation-related choices, we consider the number of observations $n$ and time-averaging length $\Delta t_{\text{avg}}$, which sets the observational noise variance $\sigma^2$.
By performing a parametric study over the three factors (Table~\ref{tab:parametricstudy}), we examine the components that formulate the posterior mean and variance (Eq.~\ref{eq:posterior_mean_variance}).

\subsection{Observations from a field measurement campaign site}
\label{sec:observations_field}

After establishing the method in the idealized setting, we apply it to field observations, where the true roughness length is unknown, to quantify the uncertainty in the inferred parameter and in predictions of unseen observations.
We use measurements from an eddy covariance flux system at the Atmospheric Radiation Measurement (ARM) Southern Great Plains (SGP) Atmospheric Observatory, at the central facility site near Lamont, Oklahoma.
Established in 1992 as the first field measurement site of the U.S. Department of Energy's ARM program, the SGP site maintains one of the longest continuous records of surface-layer turbulence measurements available \citep{stokes1994atmospheric}.
The site is selected for its relatively flat, homogeneous surface, which is broadly consistent with the horizontal-homogeneity assumption underlying MOST.
Furthermore, the site's technical report \citep{krishnamurthy2020boundary} from Pacific Northwest National Laboratory provides independent roughness length estimates against which we can compare our inferred values.
The surrounding land cover consists predominantly of agricultural land, so the effective surface roughness varies both seasonally---as a function of crop growth cycles in neighboring fields---and directionally, as a function of the upstream surface.

\begin{figure}
    \centering
    \includegraphics[width=0.6\linewidth]{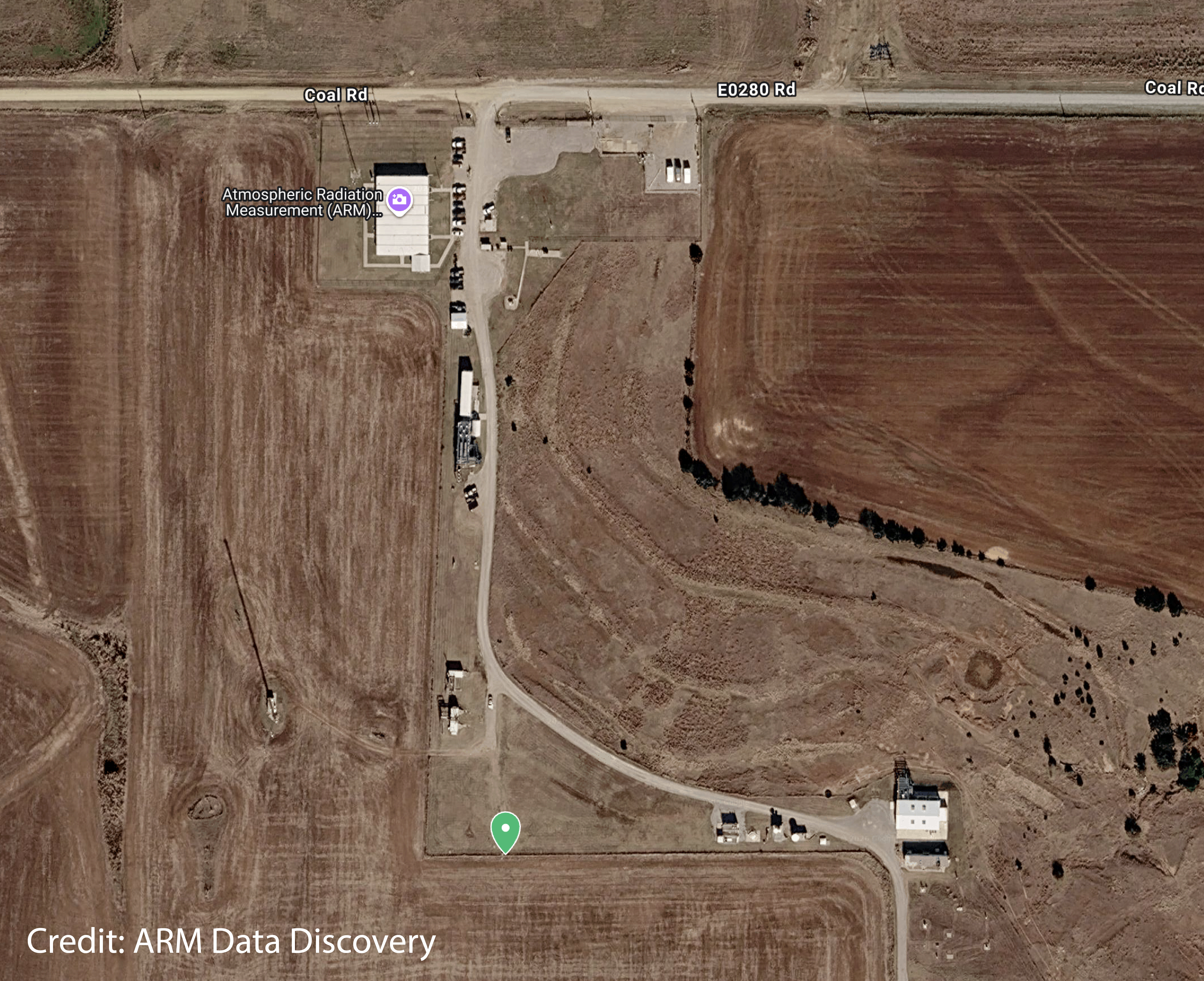}
    \caption{Aerial view of the Atmospheric Radiation Measurement (ARM) Southern Great Plains (SGP) site. The central facility (facility code: C1) of the ARM SGP site is shown in the upper-left. The eddy covariance flux system (facility code: E14) from which our measurements are taken is indicated by the green marker.
    Image from Google Earth via ARM Data Discovery.}
    \label{fig:sgp_site_google_earth}
\end{figure}

Figure~\ref{fig:sgp_site_google_earth} shows an aerial view of the central facility site of the ARM SGP site.
The central facility site (facility code: C1) is shown in the upper-left, while the eddy covariance flux system (facility code: E14) from which our measurements are taken is indicated by the green marker.
The eddy covariance flux system consists of a Gill WindMaster Pro ultrasonic anemometer and a LI-COR LI-7500 open-path gas analyzer, mounted on a small tower at a height of $3~\mathrm{m}$ above ground level \citep{osti_3029654}.
Turbulent velocity and scalar measurements are sampled at $10~\mathrm{Hz}$ and averaged over 30-minute intervals, following standard eddy covariance processing practice \citep{osti_3029654}.
Roughness length is inferred from observations spanning 2003--2014.
Following \citet{krishnamurthy2020boundary}, we retain only near-neutral observations, defined by $|L_o| > 500~\mathrm{m}$, for the inference process so that the inference is not confounded by stratification effects. 
Approximately $16\,000$ observations remain after filtering.
Predictions are made for 2018, however, under non-neutral conditions as well, since MOST extends the inferred $z_0$ to non-neutral conditions through the stability correction.

% RESULTS
\section{Results}
\label{sec:results}

The following section presents results on the probabilistic inference of surface parameters for MOST, applied first to observations sampled from LES-generated idealized ABL flows and then to field observations from the ARM SGP site.

\subsection{Inference from large-eddy simulations}
\label{sec:inference_les}

% Effect of time averaging
\begin{figure}
    \centering
    \includegraphics[width=1.0\linewidth]{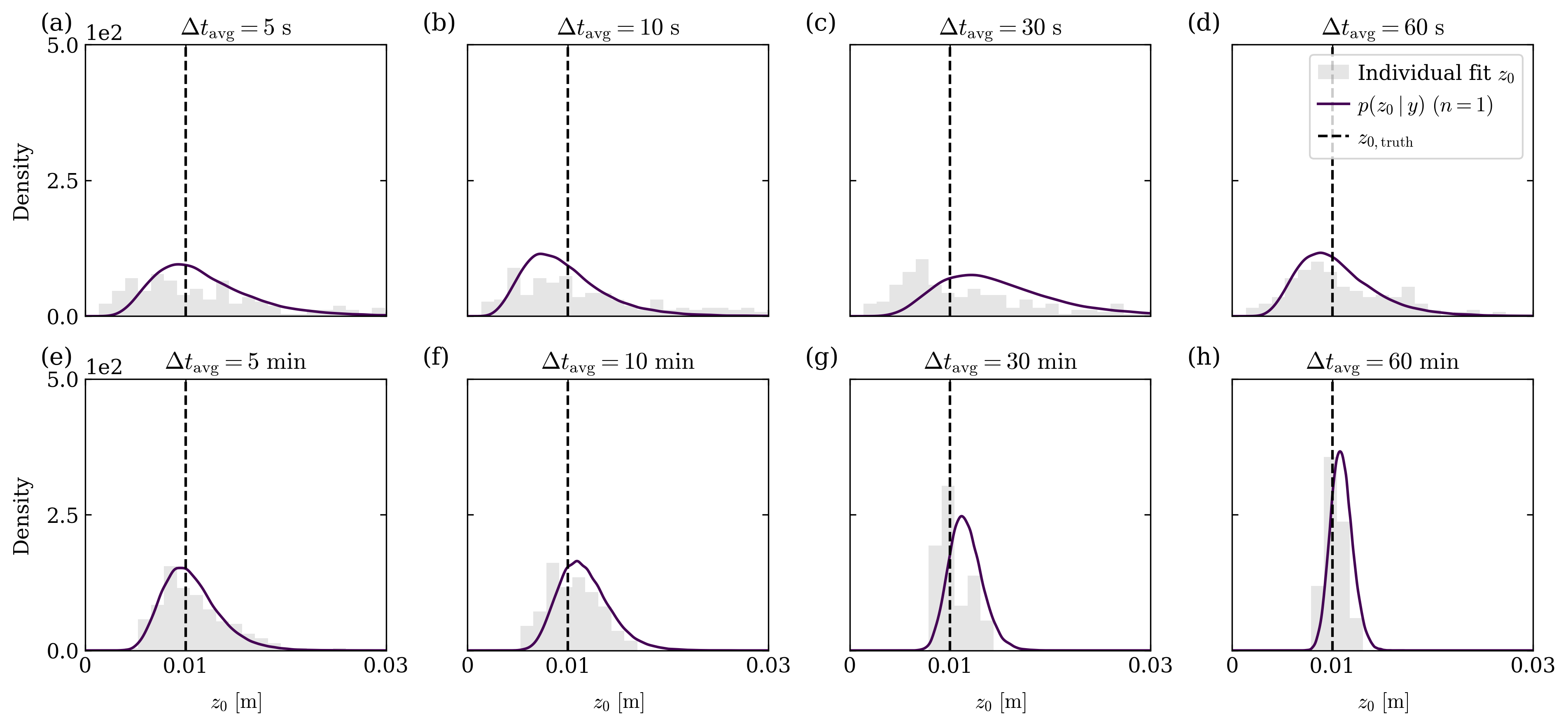}
    \caption{Posterior distributions of the roughness length inferred from a single observation ($n = 1$) for time-averaging lengths of (a) 5 s, (b) 10 s, (c) 30 s, (d) 60 s, (e) 5 min, (f) 10 min, (g) 30 min, and (h) 60 min.
    Synthetic observations are generated using LES with a true roughness length of $z_0 = 0.01~\mathrm{m}$ (dashed vertical line).
    All cases use the lognormal prior with $\mu_0=-4.73$ and $\tau_0=0.5$ (prior mean $z_0=0.01~\mathrm{m}$).
    In each panel, a histogram of the individually fit $z_0$ values at that time-averaging length is overlaid.
    }
    \label{fig:les_z0_0p01_difftavgs_singleobs_diffprior}
\end{figure}

The aerodynamic roughness length is inferred using the MOST-based wind profile model based on observations sampled from LES-generated conventionally neutral boundary layer flows.
We infer $z_0$ in isolation and exclude the displacement distance $d$, since incorporating $d$ in LES wall models has been reported to cause an unintended translational shift of the logarithmic profile rather than modify its profile curvature below the logarithmic region \citep{shin2025addressing}.
Because $z_0$ is prescribed in the wall model but never observed by the inference, recovering it from flow statistics tests the method against a known answer.
Also, because of the horizontally homogeneous surface, periodic boundary conditions, and constant geostrophic forcing, the flow statistics are quasi-stationary and independent of $xy$-location.
This means that the observational noise $\sigma^2$ arises only from finite-time sampling and is controlled by $\Delta t_{\text{avg}}$.
In this controlled setting, we explore the competing influences of the prior and observations on the posterior, before applying the method to field observations where unsteady flow and heterogeneous surface conditions also contribute to the observational noise.

We first isolate the effect of the time-averaging length.
To remove a potentially biasing effect of the prior, we place the prior mean at the truth roughness length $z_0 = 0.01~\mathrm{m}$.
Figure~\ref{fig:les_z0_0p01_difftavgs_singleobs_diffprior} shows the inferred $z_0$ posteriors obtained from a single observation across the different averaging lengths $\Delta t_{\text{avg}}$.
As the time-averaging length increases, the $z_0$ posterior narrows.
For each averaging length, we compare the inferred $z_0$ posterior with the histogram of $z_{0,i}$ obtained by fitting individual time-averaged observations.
This single-observation setup targets the estimation uncertainty of $z_0$ from a single realization, which is what the spread of individually fit $z_{0,i}$ represents empirically, and with substantial noise, as a finite-sample distribution.
The close agreement in every case establishes proof-of-concept of our method, since it accurately recovers the estimation uncertainty in $z_0$ from the $\sigma^2$ propagated through the likelihood.
However, this agreement---the recovered uncertainty matching the finite-sample distribution of individually fit $z_0$---holds specifically in the single-observation case with no prior bias.
As explained in Sect.~\ref{sec:probabilisticinference} on the balance of prior and data precisions (Eq.~\ref{eq:precisionform}), at $n=1$, the likelihood precision $n/\kappa^2\sigma^2$ can be small relative to the prior precision $1/\tau_0^2$, which then pulls the posterior to the prior mean.
The same equation implies that, for a fixed time-averaging length, increasing the number of observations $n$ both draws the posterior mean toward the truth $z_0$ and narrows the posterior.
We therefore turn to inference from multiple observations in LES, to investigate how consistently the posterior recovers the prescribed $z_0$ across the number of observations and time-averaging lengths, even with a significantly displaced prior.

Before inferring from multiple observations, we clarify a choice in the order of operations that is critical to this inference setting.
Applying MOST to the $i$-th time-averaged neutral observation gives
$
    \langle U \rangle_{i}/\langle u_* \rangle_i
    = 1/\kappa  \log\!\left( z/z_{0,i} \right),
$
whereas averaging over observations before applying MOST gives
$
    \operatorname{E} \big[ \langle U \rangle_{i} \big\rangle / \langle u_* \rangle_i \big]
    = 1/\kappa  \log\!\left( z / z_0 \right),
$
where $\langle \cdot \rangle$ denotes time averaging here, under the assumption of ergodicity, and $\operatorname{E}[\cdot]$ denotes ensemble averaging over observations.
The first procedure fits each observation separately, yielding $z_{0,i}$ for each $i$, and averages the $z_{0,i}$'s.
The second procedure averages the observations first and fits a single $z_0$ to the mean.
Neither is obviously preferred, since each observation is time-averaged and both are therefore consistent with the log law.
However, these two procedures do not result in an equivalently averaged $z_0$, and we recommend use of the second procedure for the reasons that follow.
Because the log law contains the displacement $\log z_{0}$ in its formulation, averaging the normalized profiles averages $\log z_{0}$ instead of $z_0$.
The mean profile is therefore a log law with roughness $\exp\langle \log z_{0,i} \rangle$, which is a geometric mean by construct.
By Jensen's inequality, $\exp\!\left\langle \log z_{0,i} \right\rangle \le \left\langle z_{0,i} \right\rangle$, where the latter results from the first procedure of fitting each observation separately.
The first procedure returns an arithmetic mean that will be biased upward relative to the geometric mean because it is disproportionately sensitive to outliers.
The second procedure is also chosen in the spatial aggregation of roughness over heterogeneous surfaces, where an effective $z_0$ representing multiple patches of surface types likewise requires averaging over $\log z_0$ \citep{taylor1969wind,mason1988formation}.

% Effect of prior, number of observations, and time averaging
\begin{figure}
    \centering
    \includegraphics[width=1.0\linewidth]{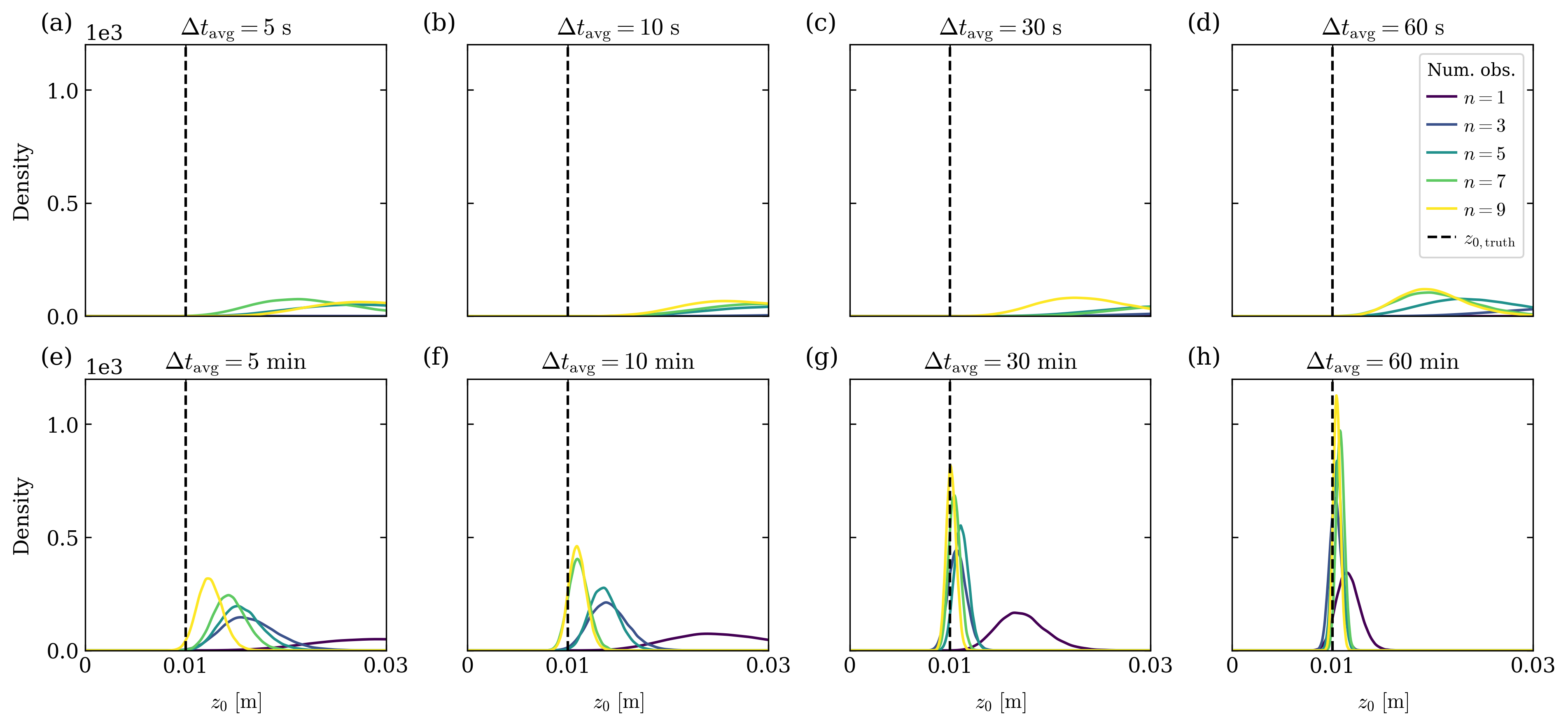}
    \caption{Posterior distributions of the roughness length inferred from $n = 1, 3, 5, 7, 9$ observations for time-averaging lengths of (a) 5 s, (b) 10 s, (c) 30 s, (d) 60 s, (e) 5 min, (f) 10 min, (g) 30 min, and (h) 60 min.
    Synthetic observations are generated using LES with a true roughness length of $z_0 = 0.01~\mathrm{m}$ (dashed vertical line).
    All cases use the lognormal prior with $\mu_0=0$ and $\tau_0=0.5$ (prior mean $z_0=1.13~\mathrm{m}$).    
    }
    \label{fig:les_z0_0p01_difftavgs_diffnumobs}
\end{figure}

% Quantified effect of prior, number of observations, and time averaging
\begin{figure}
    \centering
    \includegraphics[width=1.0\linewidth]{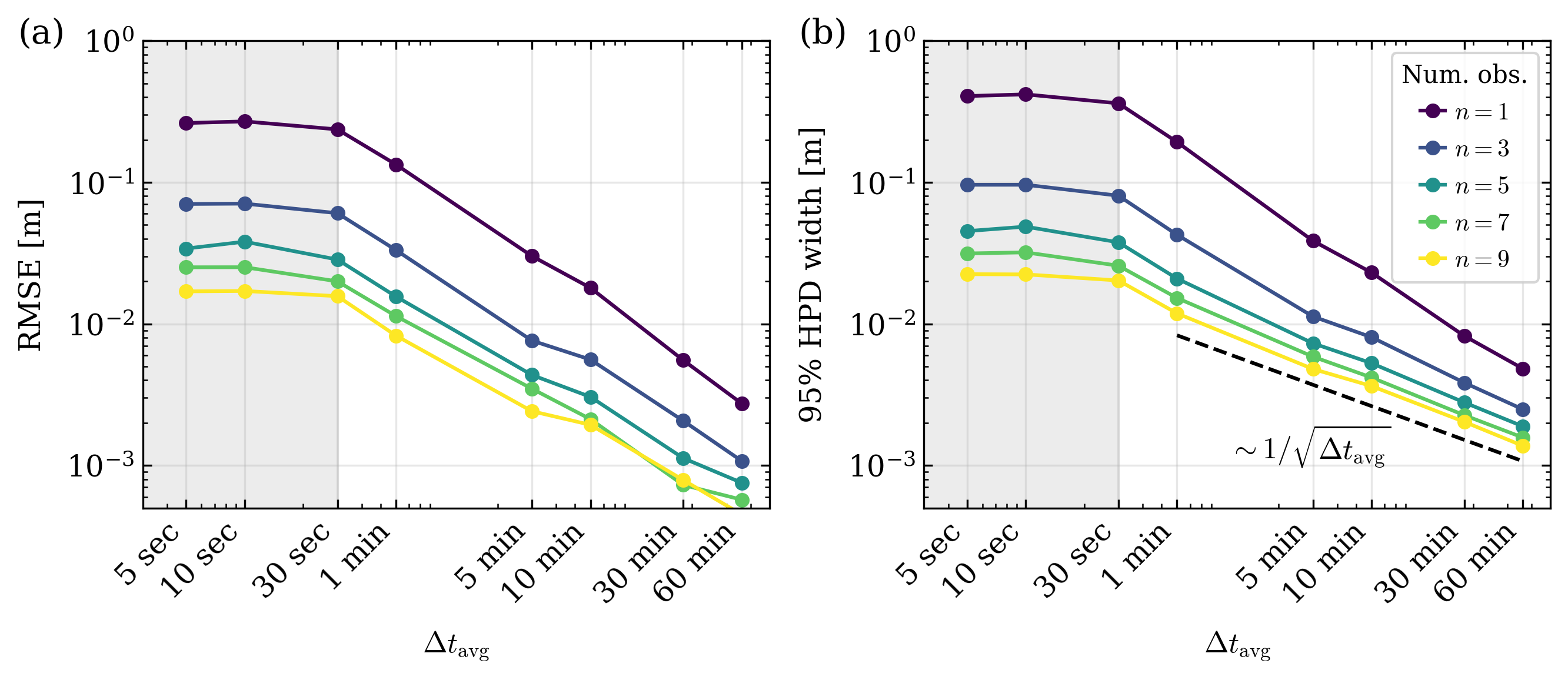}
    \caption{
    Accuracy and uncertainty of the inferred $z_0$ as functions of the time-averaging length $\Delta t_{\mathrm{avg}}$, for $n = 1, 3, 5, 7, 9$ observations.
    (a) Root mean squared error (RMSE) of the posterior mean relative to the prescribed $z_0 = 0.01~\mathrm{m}$, and (b) mean $95\%$ highest posterior density (HPD) width, both computed over $50$ independent realizations.
    The shaded region marks $\Delta t_{\mathrm{avg}} \le 30~\mathrm{s}$. 
    The dashed line in (b) indicates $1/\sqrt{\Delta t_{\mathrm{avg}}}$ scaling at fixed $n$.
    All cases use the lognormal prior with $\mu_0 = 0$ and $\tau_0 = 0.5$ (prior mean $z_0 = 1.13~\mathrm{m}$).
    }
    \label{fig:les_z0_0p01_difftavgs_diffnumobs_rmse_95hpd_ensemble}
\end{figure}

Proceeding with the second procedure, we investigate the scenario where the prior's mean is biased relative to the true value of $z_0$, while varying the number and time-averaging lengths of observations.
The prior mean is placed at $z_0 = 1.13~\mathrm{m}$, displaced upward by two orders of magnitude from the truth $z_0=0.01~\mathrm{m}$.
We infer posterior distributions from $n = 1, 3, 5, 7, 9$ observations at time-averaging lengths from 5 seconds to 60 minutes (Fig.~\ref{fig:les_z0_0p01_difftavgs_diffnumobs}).
For $\Delta t_{\text{avg}} \le 5~\mathrm{min}$ (Fig.~\ref{fig:les_z0_0p01_difftavgs_diffnumobs}a--d), the truth $z_0$ is observed to lie outside the $95\%$ highest posterior density (HPD) width at every $n$, although increasing $n$ from 1 to 9 narrows the posteriors.
These time-averaging lengths fail to average over the energy-containing turbulent eddy scales, so individual observations depart significantly from the ensemble mean, resulting in a large $\sigma^2$.
This departure observed increasingly in the posteriors inferred from shorter time-averaged observations are to the right of the truth $z_0$.
This occurs because the small likelihood precision $n/\kappa^2\sigma^2$ is small relative to the prior precision $1/\tau_0^2$, which then causes the posterior to be drawn toward the prior mean two orders of magnitude above the truth $z_0$.
This uninformative prior with a substantially displaced mean for $z_0$ is useful because it mirrors realistic inference scenarios, where the ground truth is rarely known.
This prior setup is useful in testing the sufficiency in time averaging than a prior centered on the truth $z_0$.
A prior centered on the truth $z_0$ would yield a posterior that covers the truth value regardless of the averaging, whereas here the posterior covers it only when the averaging is adequate.
Generally, the posteriors in Fig.~\ref{fig:les_z0_0p01_difftavgs_diffnumobs} narrow as both $n$ and $\Delta t_{\text{avg}}$ increase, while the influence of the prior on the posterior---manifest as a departure to the right---increases as $n$ and $\Delta t_{\text{avg}}$ decrease.
In Fig.~\ref{fig:les_z0_0p01_difftavgs_diffnumobs}f--h, we observe the sufficient time-averaging length to be 10 minutes with $n \ge 7$, 30 minutes with $n \ge 3$, and 60 minutes with $n \ge 1$, for which the posterior covers the truth $z_0$ within the $95\%$ HPD width.

To further elucidate the relationship between the prior, $n$ and $\Delta t_{\text{avg}}$, we quantify the accuracy of the posterior mean using the root mean squared error (RMSE) and the posterior uncertainty using the mean $95\%$ HPD width, both over 50 independent realizations.
Figure~\ref{fig:les_z0_0p01_difftavgs_diffnumobs_rmse_95hpd_ensemble} shows the RMSE and $95\%$ HPD width at each combination of $n$ and $\Delta t_{\text{avg}}$.
Both metrics are approximately flat for $\Delta t_{\mathrm{avg}} \leq 30~\mathrm{s}$, shaded gray in the figure.
However, both decrease monotonically for $\Delta t_{\mathrm{avg}} \geq 1~\mathrm{min}$, where the RMSE falls to a few percent of the truth $z_0$ at the longest $\Delta t_{\mathrm{avg}}$ and largest $n$ shown.
We observe that the change in behavior between $30~\mathrm{s}$ and $1~\mathrm{min}$ averaging coincides with the eddy turnover time $z/u_* \approx 45~\mathrm{s}$ at the sampled height, which sets the characteristic timescale over which the wall-bounded eddies decorrelate.
For $\Delta t_{\mathrm{avg}} \geq 1~\mathrm{min}$ the $95\%$ HPD width follows the $1/\sqrt{\Delta t_{\mathrm{avg}}}$ scaling expected once the time-averaging length spans many eddy turnover times, when the sampling variance falls inversely with the time-averaging length \citep{lenschow1994long}.
Although not shown here, the coverage of the $95\%$ HPD width, evaluated over 50 seeds, shows consistent results with the sufficient time-averaging lengths discussed above.

As expected from Eq.~\ref{eq:posterior_mean_variance}, the narrowest posterior is at $n = 9$ and $\Delta t_{\text{avg}} = 60~\mathrm{min}$, the combination of the largest $n$ and smallest $\sigma^2$ we consider.
However, in general, $n$ and $\sigma^2$ are not freely interchangeable as we observe with field observations in the next section.
Here $n$ and $\Delta t_{\text{avg}}$ are varied independently because the observations are generated by LES, so an arbitrarily long record can be simulated for any choice of averaging window given a quasi-stationary flow.
In practice, a record of fixed duration yields fewer independent observations as the window lengthens, so reducing $\sigma^2$ comes at the cost of $n$, and the ratio $n/\sigma^2$ should be targeted for improvement.
Additionally, the appropriate time-averaging length is bounded physically from below and from above.
As shown by results above (Fig~\ref{fig:les_z0_0p01_difftavgs_diffnumobs}), it is bounded from below by the requirement that the length average over several integral timescales of the energy-containing eddies, since shorter windows will fail to average over individual coherent structures.
It is bounded from above by the stationarity assumption underlying MOST, since time-averaging lengths long enough to span diurnal changes in atmospheric stability or mesoscale forcing violate the equilibrium assumption.
The conventional 30-minute window in eddy covariance practice sits between these bounds \citep{kaimal1994atmospheric,foken1996tools,osti_3029654}, and the present results support it.

The LES-based inference results illustrate the competition between the prior and observations.
When the observations are sparse or noisy, the posterior retains much of the prior information.
Consistent with the theoretical analysis in Sect.~\ref{sec:probabilisticinference}, we show that an uninformed prior can introduce bias in the inferred $z_0$, especially when inferring from insufficiently time-averaged, turbulence-resolved observations.
On the other hand, a well-informed prior can be a source of additional information that constrains the posterior beyond what least-squares regression can achieve.
This is demonstrated in the next section, where we infer $z_0$ from field observations and empirically show that a well-reasoned prior yields a posterior outperforming the least-squares fit in out-of-sample prediction in conditions of data sparsity.

\subsection{Inference from field observations}
\label{sec:inference_field}

The aerodynamic roughness length is inferred from measurements at the ARM SGP Atmospheric Observatory spanning years 2003--2014 using the MOST-based wind profile model.
In the field setting, the finite time-averaging length needed to capture the energetic scales unavoidably samples variation in the flow and upwind fetch.
Because the individual finite-time-averaged observations span a range of flow and surface conditions, the relevant quantity becomes an effective roughness length $z_0$ that reproduces the aggregate observed conditions.
Probabilistic inference from field observations yields a $z_0$ posterior centered on the value that best explains the aggregate conditions sampled by the collective observations.

We first compare the resulting $z_0$ posteriors against the roughness length estimates reported in \citet{krishnamurthy2020boundary}.
We then compare three binning or conditioning schemes---by month, by wind direction, and by their combination---and describe how each changes the inferred $z_0$ distributions.
The conditioning process is important because any variability in the observations that is not accounted for by a conditioning variable is absorbed into the observational noise variance $\sigma^2$, which then directly affects the inferred uncertainty in $z_0$.
At the ARM SGP site, conditioning by wind direction and month are physically motivated.
First, the surrounding landscape is predominantly agricultural with upstream fetches of different vegetation types by direction, so the surface roughness encountered by the flow varies directionally.
Conditioning on wind direction therefore allows the inferred $z_0$ to reflect the directionally varying surface morphology of the site.
Second, the ARM SGP site experiences changes in its agricultural landscape over a seasonal cycle, as a function of crop growth cycles in neighboring fields.
Conditioning by months ensures that the inferred $z_0$ is a function of the seasonality-driven differences.

% z0 inferred (by month) for comparison with PNNL report
\begin{figure}
    \centering
    \includegraphics[width=1.0\linewidth]{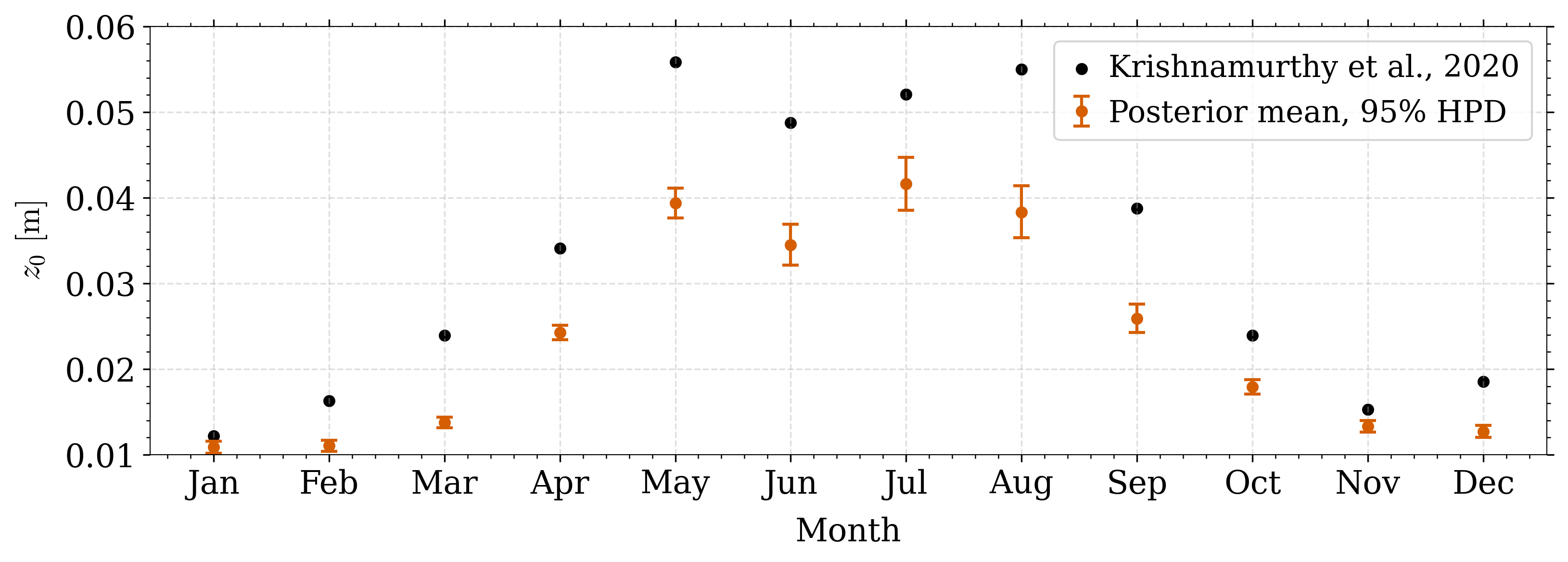}
    \caption{
    Monthly roughness length $z_0$ inferred from 30-minute-averaged near-neutral observations, binned by month and aggregated over all wind directions.
    The posterior mean and $95\%$ highest posterior density (HPD) width, shown in orange, are compared to the $z_0$ estimate from \citet{krishnamurthy2020boundary}, shown in black.
    }
    \label{fig:monthly_z0_2003-2014_30min_nn}
\end{figure}

As discussed in the previous section, $n$ and $\sigma^2$ cannot be traded freely in field observations because a record of fixed duration yields fewer independent observations as the averaging window lengthens.
Conditioning faces the same limitation, because finer conditioning can reduce $\sigma^2$ but retains fewer observations in each subset or bin, so the ratio $n/\sigma^2$ that governs the data precision does not necessarily improve.
Nonetheless finer conditioning is desirable, since gaining insights into the relationship between $z_0$ and different surface and flow conditions can spur further model development.
Therefore our inference method must reliably estimate $z_0$ and quantify the associated uncertainty in the smaller $\sigma^2$ and $n$ regime.
In this study, we aim to improve the prior by setting its mean and standard deviation in the transformed, normally distributed space based on sample statistics of all near-neutral observations from 2003--2014, 
\begin{equation}
    \mu_0 = \frac{1}{N}\sum_{i=1}^{N} \log z_{0,i} = -3.940,
    \qquad
    \tau_0 = \sqrt{ \frac{1}{N-1}\sum_{i=1}^{N} \left( \log z_{0,i} - \mu_0 \right)^2 } = 1.125.
\end{equation}
The rationale is that, in data-sparse conditions, we will fall back on the global statistics for the site, which we assume to better reflect the long-term site characteristics than a few noisy observations.
We evaluate the validity of this assumption through predictions of unseen observations.

First to compare with monthly $z_0$ estimates from \citet{krishnamurthy2020boundary}, we infer $z_0$ from 30-minute-averaged near-neutral observations of wind speed normalized by friction velocity, binned by month.
Here observations are integrated over wind direction, which yield approximately $\mathcal{O}(10^3)$ observations per monthly bin.
The posterior mean and $95\%$ HPD width of the inferred $z_0$ per month are shown in Fig.~\ref{fig:monthly_z0_2003-2014_30min_nn}. 
As expected by the large number of observations used in the inference process, the $95\%$ HPD width is very narrow.
Some discrepancy between the inferred values and the estimates from \citet{krishnamurthy2020boundary} is observed but expected, as the report relies on measurements from a 60-m meteorological tower located approximately $100~\mathrm{m}$ northwest of the eddy covariance flux system used here.
Nevertheless, the inferred $z_0$ accurately reproduces the reported seasonal trend.
During winter months (December--February), the inferred $z_0$ remains just above $0.01~\mathrm{m}$ with narrowed $95\%$ HPD widths, which reflects the smoother aerodynamic roughness of the dormant agricultural surface.
The inferred $z_0$ posteriors for the spring and summer months are both higher in value, reaching a peak  around $0.04~\mathrm{m}$ in July, and wider (approximately $\pm0.003~\mathrm{m}$ at the peak).
This seasonal pattern is attributable to the agricultural characteristics of the surrounding land cover: crop growth in the spring and summer is reflected in an increase in $z_0$, while subsequent harvest in the fall is reflected in a decrease in $z_0$ that extends into the winter.

% z0 inferred (by wind direction) for comparison with PNNL report
\begin{figure}
    \centering
    \includegraphics[width=1.0\linewidth]{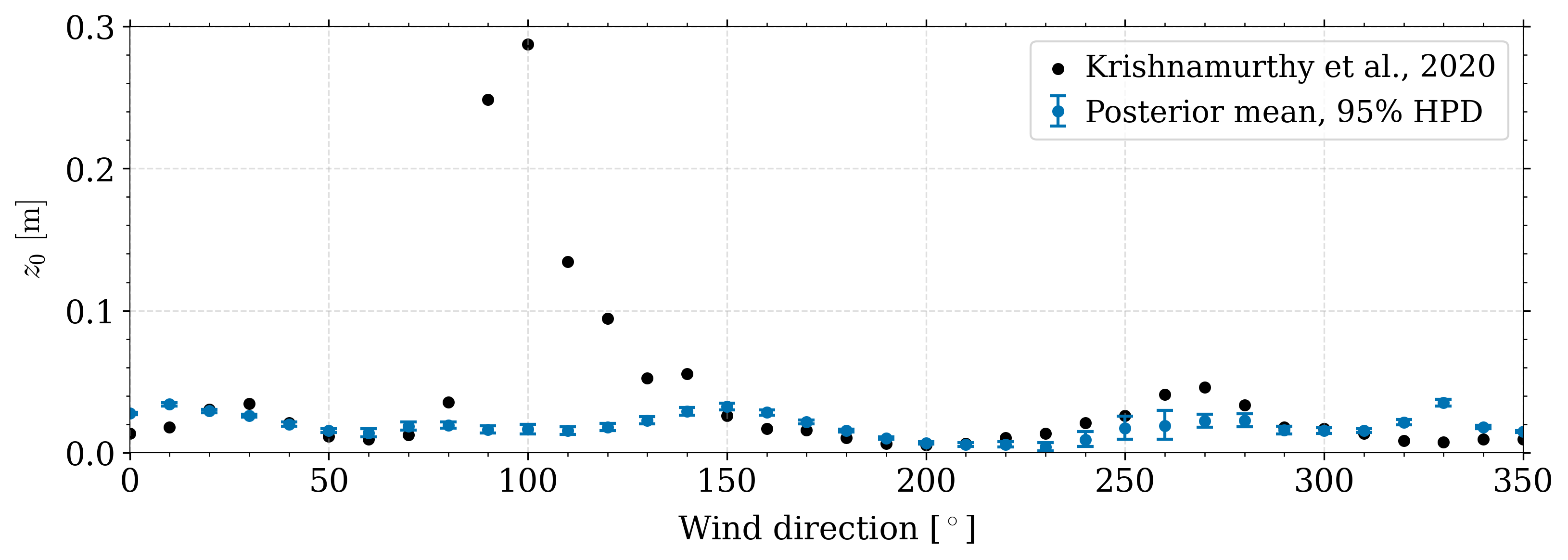}
    \caption{
    Wind-direction-based roughness length $z_0$ inferred from 30-minute-averaged near-neutral observations, binned by wind direction (centered $10^\circ$) and aggregated over all months.
    The posterior mean and $95\%$ highest posterior density (HPD) width, shown in blue, are compared to the $z_0$ estimate from \citet{krishnamurthy2020boundary}, shown in black.
    }
    \label{fig:wdir_z0_2003-2014_30min_nn}
\end{figure}

The increased $95\%$ HPD widths observed in the summer months also includes directional variability in land use surrounding the site, which is naturally captured when observations are conditioned on wind direction.
Figure~\ref{fig:wdir_z0_2003-2014_30min_nn} shows the inferred $z_0$ posterior mean and $95\%$ HPD widths binned by wind direction in centered $10^\circ$ increments.
This time, we integrate observations over month within each wind direction bin.
As noted previously, some discrepancy between the inferred values and estimates from \citet{krishnamurthy2020boundary} is expected given the differing measurement sources.
In particular, the 60-m meteorological tower used in \citet{krishnamurthy2020boundary} has maintenance buildings to its east that contribute to significantly increased $z_0$ estimates local to the easterly wind direction bins ($80$--$120^\circ$).
This phenomenon is absent in the eddy covariance flux system observations used in this study.
The source-dependent discrepancy underscores the importance of accounting for local surface heterogeneity when modeling $z_0$, as even the instrument location can substantially influence the inferred roughness in specific directions.

% z0 inferred from joint month x wind direction
\begin{figure}
    \centering
    \includegraphics[width=1.0\linewidth]{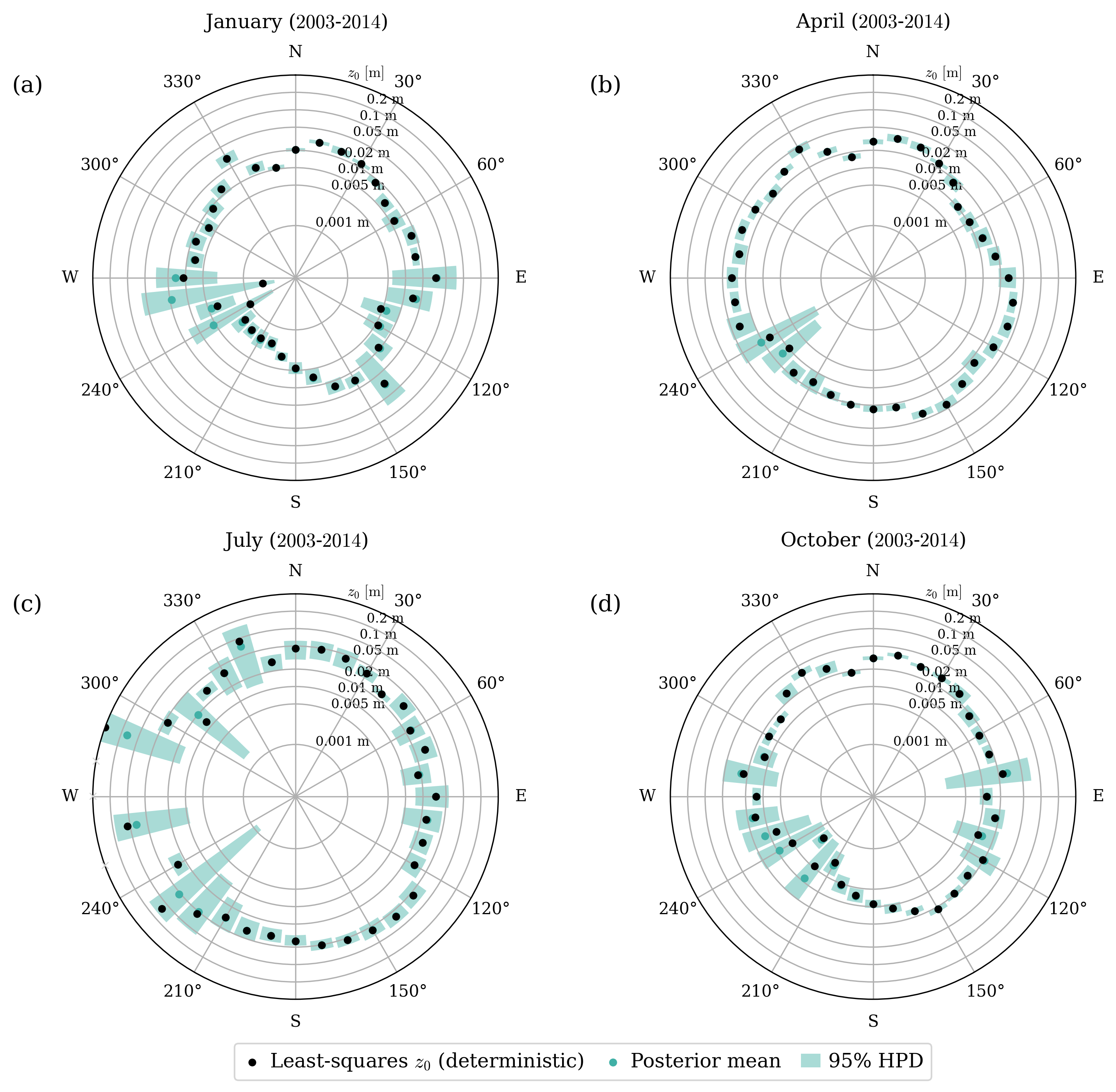}
    \caption{
    Inferred roughness length $z_0$ displayed as a roughness rose, where 30-minute-averaged near-neutral observations are binned by both month and wind direction (centered $10^\circ$).
    Representative months for each of the four seasons are shown: (a) January, (b) April, (c) July, and (d) October.
    The mean (circle) and $95\%$ highest posterior density (HPD) width (shaded region) are shown in teal for each wind direction bin, while corresponding least-squares estimates are shown as black circles.
    Radial ticks indicate the magnitude of $z_0$.
    }
    \label{fig:roughness_rose_2003-2014_30min_nn}
\end{figure}

A third conditioning scheme bins the observations jointly in month and wind direction, yielding $12 \times 36 = 432$ bins.
Here we emphasize that the aim of this study is not to identify the best conditioning approach, but to show that this choice influences the inferred $z_0$ posteriors.
Here we also introduce least-squares $z_0$ estimates under the same joint conditioning. 
These provide the fair comparison against the state-of-practice approach.
Figure~\ref{fig:roughness_rose_2003-2014_30min_nn} shows the resulting \textit{roughness rose}---inferred $z_0$ posterior mean (circle) and $95\%$ HPD width (shaded region) at each wind direction bin visualized analogously to a wind rose---for four months representative of the four seasons.
Across the four months, the seasonal trend of lower $z_0$ in winter and higher $z_0$ in summer is clearly visible along with the month-specific directional dependence.
The direction bins now show $z_0$ uncertainty that was not apparent in either the separate monthly or wind direction-based conditioning, nor in the corresponding least-squares $z_0$ estimate.
The increase in uncertainty is a direct result of the finer binning, since spreading the available near-neutral observations across 432 bins leaves few observations in each bin.
Some contain as few as $\mathcal{O}(1)$ observations.
Conditioning therefore shows the trade-off between resolving the physical dependence of $z_0$ on season and upstream fetch and retaining enough observations per bin to constrain $z_0$.
The data-sparse regimes are where we expect to see noticeable differences between posterior mean and least-squares estimates, which then result in differing wind speed predictions when put to operational use.
As the posterior distribution asymptotically goes towards the least-squares estimate in the large-data limit, we observe that in many of the bins the posterior mean and least-squares $z_0$ coincide.
However, an important benefit of the proposed Bayesian method is that the inferred surface parameters inherently capture the associated uncertainty, in contrast to the deterministic $z_0$ estimates that are unaware of the uncertainty introduced by limited, noisy observations.

\subsection{Predictions of unseen field observations}
\label{sec:predictions}

We evaluate how well observations not used in the inference can be predicted by the inferred $z_0$ posteriors, and how this compares to predictions by the least-squares estimate.
The observed or training data consists of 30-minute-averaged near-neutral observations from 2003--2014, and unseen test data consists of 30-minute-averaged observations from 2018.

The training data are restricted to near-neutral conditions while the test data span all stability conditions, since MOST extends the logarithmic profile to non-neutral conditions through the stability correction.
Using our forward model (Eq.~\ref{eq:forwardmodel}), the unseen observation to be predicted is the stability-corrected, friction-velocity-normalized wind speed
\begin{equation}
    \tilde{y}_i = \frac{\langle U \rangle_i}{\langle u_* \rangle_i}
    + \frac{1}{\kappa} \psi_m \left( \frac{z}{\langle L_o \rangle}_i \right),
    \label{eq:predictand}
\end{equation}
and $\hat{\tilde{y}}_i$ denotes its prediction.

We consider both a deterministic and a probabilistic prediction using the Bayesian inferred $z_0$. 
The deterministic prediction is the posterior predictive mean, conditioned jointly on month and wind direction, while the probabilistic prediction is the corresponding posterior predictive distribution (Eq.~\ref{eq:posteriorpredictivedistribution}).
Both are compared against the deterministic prediction from the least-squares estimate on the training data, also conditioned jointly.
This allows us to address whether the deterministic prediction from the Bayesian inferred $z_0$ improves on the least-squares approach and whether the probabilistic prediction adds value beyond it.

The accuracy of deterministic predictions is quantified with the root mean squared error and the mean absolute error over $N$ unseen observations in each conditioning bin,
\begin{equation}
    \text{RMSE} = \sqrt{\frac{1}{N}\sum_{i=1}^{N}
    \big(\tilde{y}_i - \hat{\tilde{y}}_i\big)^2}, 
    \quad 
    \text{MAE} = \frac{1}{N}\sum_{i=1}^{N}
    \big|\tilde{y}_i - \hat{\tilde{y}}_i\big|.
\end{equation}

The accuracy of probabilistic predictions is quantified with the continuous ranked probability score (CRPS), a proper scoring rule that provides a principled basis for comparing deterministic and probabilistic predictions \citep{gneiting2007strictly}.
The CRPS measures the discrepancy between the prediction distribution and the observed value as the integrated squared difference between the prediction cumulative distribution function and the step function at the observation.

The CRPS can be expressed in closed form for a Gaussian prediction.
Our posterior predictive distribution (Eq.~\ref{eq:posteriorpredictivedistribution}) is normally distributed with predictive mean $\mu_{\text{pred}} = (\log z - \mu_n)/\kappa$ and variance $\sigma_{\text{pred}}^2 = \tau_n^2/\kappa^2 + \sigma^2$.
Defining the standardized residual $w_i = (\tilde{y}_i - \mu_{\text{pred}})/\sigma_{\text{pred}}$, the CRPS of our probabilistic prediction becomes
\begin{equation}
    \mathrm{crps}_i = \sigma_{\text{pred}}\left[\, w_i\big(2\Phi(w_i) - 1\big)
    + 2\varphi(w_i) - \frac{1}{\sqrt{\pi}} \,\right],
    \label{eq:crpsgauss}
\end{equation}
where $\Phi$ and $\varphi$ denote the standard normal cumulative distribution and probability density functions, respectively.
Applying Eq.~\eqref{eq:crpsgauss} to each of the $N$ test observations $\tilde y_i$ in the given bin and averaging gives the aggregate score
\begin{equation}
    \text{CRPS} = \frac{1}{N}\sum_{i=1}^{N} \mathrm{crps}_i.
    \label{eq:crps}
\end{equation}
On the other hand, the CRPS for a deterministic prediction reduces to the absolute error.
Therefore, the aggregate CRPS for deterministic predictions of the $N$ unseen observations in each conditioning bin is simply the MAE.

\begin{figure}
    \centering
    \includegraphics[width=1.0\linewidth]{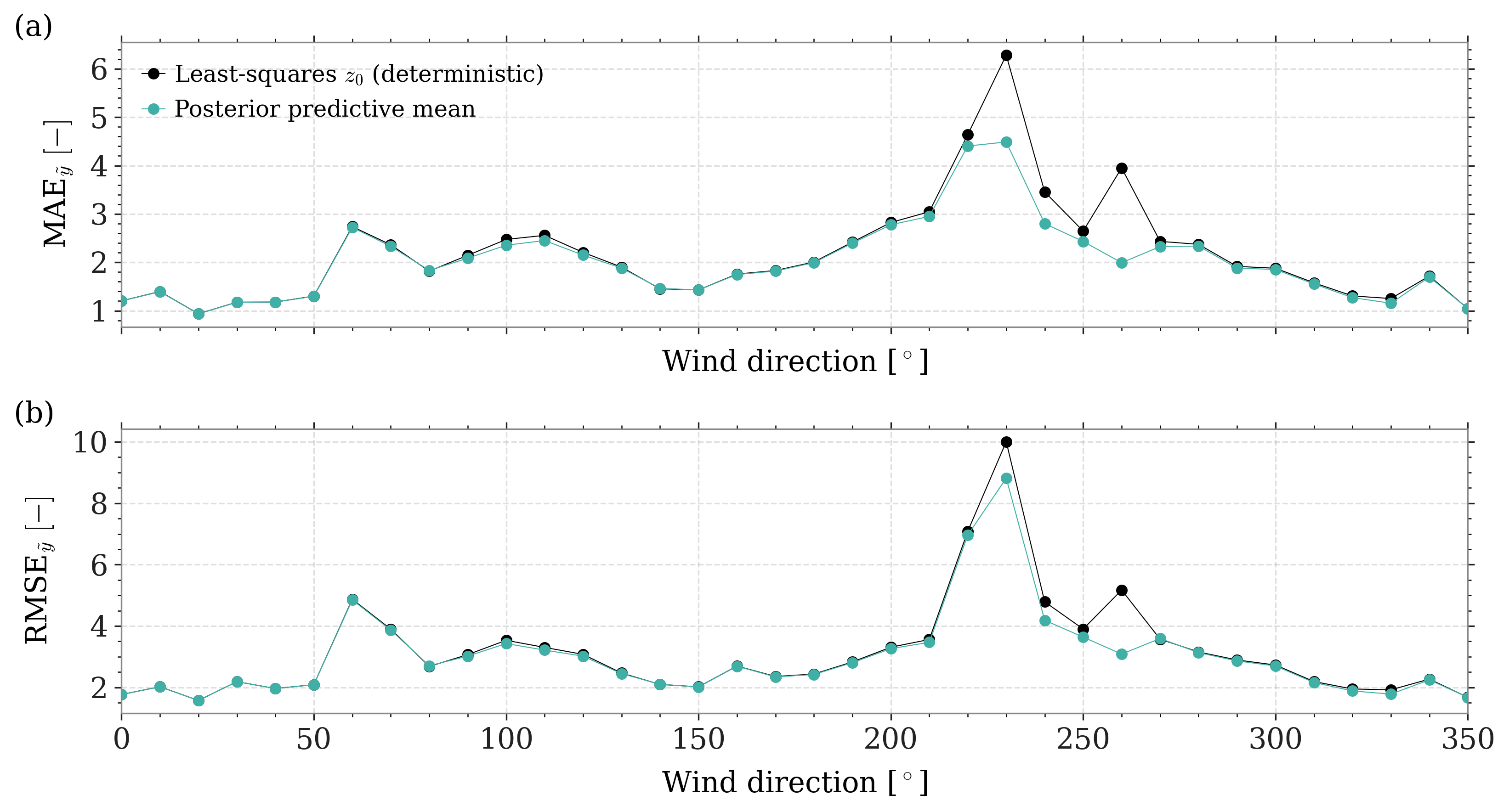}
    \caption{
    (a) Mean absolute error (MAE) and (b) root mean square error (RMSE) for deterministic predictions of unseen observations $\tilde{y}$ from 2018, as a function of wind direction.
    Predictions use $z_0$ based on 30-minute-averaged near-neutral observations from 2003--2014, conditioned jointly on month and wind direction.
    Given metric for the deterministic prediction from the least-squares $z_0$ estimate is shown in black, while for the equivalent for the posterior predictive mean is shown in teal.
    }
    \label{fig:mae_rmse_wdir_2018}
\end{figure}

We first compare the deterministic predictions---the posterior predictive mean against the prediction by the least-squares $z_0$ estimate---across wind direction bins (Fig.~\ref{fig:mae_rmse_wdir_2018}).
Predictions use the $z_0$ inferred from observations conditioned jointly on month and wind direction.
The RMSE and MAE are then averaged over months within each wind direction bin to visualize the metrics as a function of wind direction, which we find most informative.
Over most of the wind direction bins, the posterior predictive mean and prediction by the least-squares $z_0$ estimate yield indistinguishable RMSE and MAE.
This is because the posterior mean converges to the least-squares estimate with increasing $n$ at an approximately unchanging $\sigma^2$, as indicated by the asymptotic behavior of the precision-weighted average (Eq.~\ref{eq:precisionform}).
However, the two predictions significantly differ in the $210$--$270^\circ$ bins, where the posterior predictive mean consistently improves upon the prediction by the least-squares estimate.
Notably, at $230^\circ$ the posterior predictive mean reduces the MAE from 6.3 to 4.5 and the RMSE from 10.0 to 8.8.
Also, at $260^\circ$ it reduces the MAE from 4.0 to 2.0 and the RMSE from 5.2 to 3.1.
We note that these bins contain the fewest observations and correspond to the data-sparse conditions identified in the theoretical analysis in Sec.~\ref{sec:probabilisticinference}.
Here we find that the least-squares estimate is fit to the $\mathcal{O}(1)$ noisy observations in each bin, whereas the posterior predictive mean retains the prior information which we informed with the sample statistics of all near-neutral observations from 2003-2014.
The consistent improvement in deterministic-prediction metrics indicates that the prior adds value in data-sparse regime, and therefore supports the prior assumption we adopted for the site.
That is, the long-term site characteristics captured by all near-neutral observations prove useful to fall back on when only a few noisy observations are available.
By leveraging Bayes' theorem together with knowledge of surface-atmosphere interactions, we have shown how Bayesian inference can reduce predictive error relative to the state-of-practice least-squares method.

\begin{figure}
    \centering
    \includegraphics[width=1.0\linewidth]{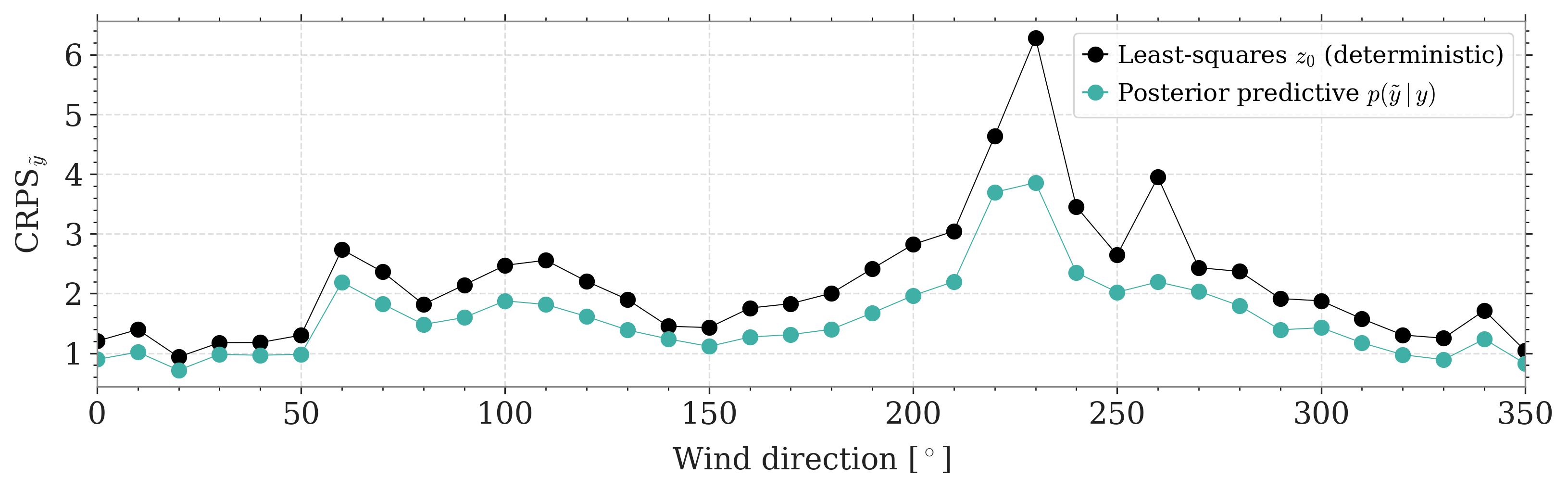}
    \caption{
    Continuous ranked probability score (CRPS) for predictions of unseen observations $\tilde{y}$ from 2018, as a function of wind direction.
    Predictions use $z_0$ based on 30-minute-averaged near-neutral observations from 2003--2014, conditioned jointly on month and wind direction.
    CRPS for the deterministic prediction from the least-squares $z_0$ estimate is shown in black, while the equivalent for the posterior predictive distribution is shown in teal.
    Lower scores are better.
    }
    \label{fig:crps_wdir_2018}
\end{figure}

Using the CRPS, we also compare the posterior predictive distribution against the deterministic prediction by the least-squares $z_0$ estimate across wind direction bins (Fig.~\ref{fig:crps_wdir_2018}).
In contrast to the deterministic-prediction results, which show improvements by the Bayesian approach only in the data-sparse bins, the probabilistic-prediction metric shows a lower score for the posterior predictive distribution relative to the prediction by the least-squares $z_0$ in every wind direction bin.
A consistent improvement of roughly $20$--$30\%$ in the CRPS is observed across all direction bins.
The improvement margin is largest in the $210$--$270^\circ$ bins, where the deterministic prediction reaches $6.3$ at $230^\circ$ against $3.9$ for the full posterior predictive.
This is because CRPS evaluates the predicted probabilistic distribution against the set of unseen observations in a given bin, and rewards a predictive spread that matches the spread of the observations. 
By contrast, a deterministic prediction has zero spread.
The CRPS therefore quantifies the predictive skill lost by ignoring the uncertainty in $z_0$ estimates.
The consistent improvement margin across every direction bin shows that the probabilistic prediction improves skill everywhere, not only in the data-sparse regime.

\section{Conclusions}
\label{sec:conclusions}

The current study demonstrates a Bayesian approach to inferring surface parameters for MOST.
For the aerodynamic roughness length inferred in isolation, the posterior and posterior predictive distributions admit closed-form expressions, so that probabilistic estimates are obtained at negligible computational cost relative to the state-of-practice deterministic method based on least-squares fits.

Applied to idealized conventionally neutral boundary layers from large-eddy simulation, the method recovered the prescribed roughness length, demonstrating proof-of-concept of the method.
By testing different prior- and observation-related choices, we observed that the prescribed $z_0$ was recovered in the $95\%$ HPD width once the combination of time-averaging length and number of observations $n$ were sufficient, which was $n \geq 7$ at $10$-minute averaging, $n \geq 3$ at $30$-minute averaging, and $n \geq 1$ at $60$-minute averaging.
We showed that the $z_0$ posterior's behavior depends increasingly on the prior as the observations become sparser and noisier, which reflects how the relative contributions are governed by the data and prior precisions.
Applied to field observations from the ARM Southern Great Plains observatory, the method yielded site-specific probabilistic estimates of $z_0$ that depend on both season and upstream fetch.
Resolving the dependence of $z_0$ on month and wind direction requires finer bins, but finer bins leave fewer observations in each, which increases the effect of the prior on the inferred $z_0$ posteriors.
By constructing the prior from sample statistics pooled over all near-neutral observations at the site, we showed that Bayesian inference of $z_0$ improves on the conventional least-squares method in both predictive accuracy and uncertainty quantification.
Predictions for unseen observations in a test year revealed that the inferred $z_0$ improved the deterministic prediction relative to the least-squares $z_0$ estimate in data-sparse regimes, and improved the probabilistic prediction by approximately $20$--$30\%$ in CRPS across all wind directions considered.

Several directions for future work follow naturally from this study. 
The closed-form posterior derived in this study inherently contains the same assumptions and limitations as MOST in general, such as horizontal homogeneity and statistical stationarity, but the Bayesian methodologies leveraged are more general.
Therefore, future work may extend the UQ methodology to alternative representations of surface-atmosphere interactions beyond MOST.
The methodology may also be extended to real-time inference from field measurement systems to enable online updating of roughness parameters for use in operational Earth system models.
While the two conditioning variables used here were selected \textit{a priori}, future work could investigate principled, data-driven methods for identifying additional conditioning variables that best characterize the effective roughness length at a given site.
A method for autonomous discovery of the relevant classification variables from the observations themselves would be highly useful for the meteorological and broader data-driven atmospheric modeling community.

\paragraph{Acknowledgments}
Simulations were performed on the Stampede3 supercomputer under the NSF ACCESS project ATM170028.

\paragraph{Funding Statement}
E.Y.S. and M.F.H. acknowledge the support of Office of Naval Research, Young Investigator Program (YIP), grant no. N000142512045.

\paragraph{Conflict of Interest}
The authors declare there are no conflicts of interest for this manuscript.

\paragraph{Data availability}
The field observations analyzed in this study can be obtained through ARM Data Discovery (\url{https://adc.arm.gov/discovery}).
The relevant datastream is sgp30ecorE14.b1, accessible at \url{https://doi.org/10.5439/1025039}.
The code and LES data that support the findings of this study will be made openly available upon publication.

\begin{appendix}
\crefalias{section}{appendix}

\section{Joint inference of surface parameters}
\label{sec:appendix2}

We detail how multiple surface parameters may be jointly inferred from wind speed observations at several vertical levels.
We formulate a forward model for inferring the roughness length $z_0$, the displacement distance $d$, and the empirically tuned coefficients $\beta$ and $\gamma$ in the non-integrated functions for the stable and unstable regimes, respectively.
This forward model maps parameters $\theta = (z_0, d, \beta, \gamma)^{T}$ to the friction-velocity-normalized wind speed of observation $i = 1, \dots, n$ at height $z_k$, $k = 1, \dots, r$,
\begin{equation}
    f_{ik}(\theta) = \frac{1}{\kappa} \left[ \log\!\left(\frac{z_k - d}{z_0}\right) - \psi_m \left( \frac{z_k}{ \langle L_o \rangle_i },\beta,\gamma\right)  \right],
    \label{eq:forwardmodel2}
\end{equation}
where the heights $z_k$ and the mean Obukhov length $\langle L_o \rangle_i$ are known.
The observations are related to the model prediction $f(\theta)$ through
\begin{equation}
  y = f(\theta) + \eta, \qquad \eta \sim \mathcal{N}\!\left(0, \Sigma_y \right),
  \label{eq:obsmodel}
\end{equation}
where $\Sigma_y$ is the observational error covariance matrix.
The associated log likelihood is
\begin{equation}
  \log p(y \mid \theta) = -\frac{1}{2}
    \left\| y - f(\theta) \right\|^{2}_{\Sigma_y}
    - \frac{1}{2} \log \left| \Sigma_y \right|
    - \frac{nr}{2} \log (2\pi),
  \label{eq:loglik}
\end{equation}
where $\|\cdot\|_{A}$ is the Mahalanobis norm.

The roughness length and the displacement distance are strictly positive but vary over orders of magnitude across land cover types \citep{wiernga1993representative}.
The displacement distance is further constrained above by the lowest measurement level.
The stability coefficients are bounded but otherwise poorly constrained, since reported values differ substantially between field campaigns \citep{businger1971flux,dyer1974review}.
Because the parameters are subject to distinct constraints, we choose the prior distributions accordingly
\begin{align}
  p(z_0) &= \mathrm{lognormal}( z_0 \mid \mu_{z_0}, \tau_{z_0}^{2}), \\
  p(d) &= \mathrm{lognormal}(d \mid \mu_d, \tau_d^{2}), \\
  p(\beta) &= \mathcal{U}(\beta \mid a_\beta, b_\beta), \\
  p(\gamma) &= \mathcal{U}(\gamma \mid a_\gamma, b_\gamma).
  \label{eq:priors}
\end{align}
We choose the parameters to be independent \textit{a priori} because of the absence of a credible joint prior.
However, we note the independence assumption constrains only the priors, and correlations can arise in the posterior through the coupling of the parameters in the forward model.
In particular, we expect a negative correlation between $z_0$ and $d$ \citep{macdonald1998improved,shin2024multi}.

Sampling is performed in the transformed, unconstrained space.
We introduce $\xi = (\xi_{z_0}, \xi_d, \xi_\beta, \xi_\gamma)^{T}$ through the logarithmic map for each length parameter, $j \in \{z_0, d\}$,
\begin{equation}
  \xi_{z_0} = \log z_0 \iff z_0 = e^{\xi_{z_0}},
  \qquad
  \xi_d = \log d \iff d = e^{\xi_d},
  \label{eq:logmap}
\end{equation}
and the logit map for each bounded coefficient $c \in \{\beta, \gamma\}$,
\begin{equation}
  \xi_c = \log \left( \frac{c - a_c}{b_c - c} \right)
  \iff
  c = a_c + (b_c - a_c)\, S(\xi_c),
  \qquad
  S(\xi) = \left( 1 + e^{-\xi} \right)^{-1}.
  \label{eq:logitmap}
\end{equation}
Each lognormal prior is mapped to a normally distributed prior, 
\begin{equation}
  \log \left[ p(e^{\xi_j}) \left| \frac{\mathrm{d}j}{\mathrm{d}\xi_j} \right| \right]
  = -\frac{(\xi_j - \mu_j)^{2}}{2\tau_j^{2}} + \mathrm{const},
  \label{eq:priorphi}
\end{equation}
while each uniform prior is mapped to a standard logistic prior,
\begin{equation}
  \log \left[ p(c) \left| \frac{\mathrm{d}c}{\mathrm{d}\xi_c} \right| \right]
  = -\xi_c - 2 \log \left( 1 + e^{-\xi_c} \right).
  \label{eq:priorpsi}
\end{equation}

Combining the log likelihood (Eq.~\ref{eq:loglik}) and the priors (Eq.~\ref{eq:priorphi} and Eq.~\ref{eq:priorpsi}), the log posterior in the unconstrained space is
\begin{equation}
  \log p(\xi \mid y)
  = -\frac{1}{2}
    \left\| y - f\!\left( \theta(\xi) \right) \right\|^{2}_{\Sigma_y}
    - \sum_{j \in \{z_0, d\}} \frac{(\xi_j - \mu_j)^{2}}{2\tau_j^{2}}
    - \sum_{c \in \{\beta, \gamma\}}
      \left[ \xi_c + 2 \log \left( 1 + e^{-\xi_c} \right) \right]
    + \mathrm{const},
  \label{eq:logpost}
\end{equation}
subject to $\xi_d < \log \left( \min_k z_k \right)$, since the displacement distance must be below the lowest measurement level.
Equation~\ref{eq:logpost} can be sampled using a random-walk Metropolis--Hastings algorithm with a normal proposal.
The samples can then be mapped back to the constrained, physical space through Eq.~\ref{eq:logmap} and Eq.~\ref{eq:logitmap}.

\end{appendix}

\bibliography{references_}

@article{golbazi2019methods,
  title={Methods to estimate surface roughness length for offshore wind energy},
  author={Golbazi, Maryam and Archer, Cristina L},
  journal={Advances in Meteorology},
  volume={2019},
  number={1},
  pages={5695481},
  year={2019},
  publisher={Wiley Online Library},
  doi = {10.1155/2019/5695481}
}

@article{martano2000estimation,
  title={Estimation of surface roughness length and displacement height from single-level sonic anemometer data},
  author={Martano, Paolo},
  journal={Journal of Applied Meteorology},
  volume={39},
  number={5},
  pages={708--715},
  year={2000},
  doi = {10.1175/1520-0450(2000)039<0708:EOSRLA>2.0.CO;2}
}

@article{prigent2005estimation,
  title={Estimation of the aerodynamic roughness length in arid and semi-arid regions over the globe with the ERS scatterometer},
  author={Prigent, Catherine and Tegen, Ina and Aires, Filipe and Marticorena, B{\'e}atrice and Zribi, Merhez},
  journal={Journal of Geophysical Research: Atmospheres},
  volume={110},
  number={D9},
  year={2005},
  publisher={Wiley Online Library},
  doi = {10.1029/2004JD005370}
}

@inproceedings{shin2024multi,
  title={Multi-fidelity modeling and uncertainty quantification of heterogeneous roughness},
  author={Shin, Ethan Y and Chan, Miles and Wang, Jianyu and Zahtila, Tony and Gorle, Catherine and Iaccarino, Gianluca and Howland, Michael},
  booktitle={Center for Turbulence Research Proceedings of the Summer Program},
  year={2024},
}

@article{macdonald1998improved,
  title={An improved method for the estimation of surface roughness of obstacle arrays},
  author={Macdonald, RW and Griffiths, RF and Hall, DJ},
  journal={Atmospheric environment},
  volume={32},
  number={11},
  pages={1857--1864},
  year={1998},
  publisher={Elsevier},
  doi = {10.1016/S1352-2310(97)00403-2}
}

@article{bou2007parameterization,
  title={On the parameterization of surface roughness at regional scales},
  author={Bou-Zeid, Elie and Parlange, Marc B and Meneveau, Charles},
  journal={Journal of the atmospheric sciences},
  volume={64},
  number={1},
  pages={216--227},
  year={2007},
  doi = {10.1175/JAS3826.1}
}

@article{bergeron1992estimating,
  title={Estimating shear velocity and roughness length from velocity profiles},
  author={Bergeron, Normand E and Abrahams, Athol D},
  journal={Water Resources Research},
  volume={28},
  number={8},
  pages={2155--2158},
  year={1992},
  publisher={Wiley Online Library},
  doi = {10.1029/92WR00897}
}

@article{floors2021satellite,
  title={Satellite-based estimation of roughness lengths and displacement heights for wind resource modelling},
  author={Floors, Rogier and Badger, Merete and Troen, Ib and Grogan, Kenneth and Permien, Finn-Hendrik},
  journal={Wind Energy Science},
  volume={6},
  number={6},
  pages={1379--1400},
  year={2021},
  publisher={Copernicus GmbH},
  doi = {10.5194/wes-6-1379-2021}
}

@article{hammond2012roughness,
  title={Roughness length estimation along road transects using airborne LIDAR data},
  author={Hammond, DS and Chapman, L and Thornes, JE},
  journal={Meteorological Applications},
  volume={19},
  number={4},
  pages={420--426},
  year={2012},
  publisher={Wiley Online Library},
  doi = {10.1002/met.273}
}

@article{he2017estimation,
  title={Estimation of roughness length at Hong Kong International Airport via different micrometeorological methods},
  author={He, YC and Chan, PW and Li, QS},
  journal={Journal of Wind Engineering and Industrial Aerodynamics},
  volume={171},
  pages={121--136},
  year={2017},
  publisher={Elsevier},
  doi = {10.1016/j.jweia.2017.09.019}
}

@book{stull2012introduction,
  title={An introduction to boundary layer meteorology},
  author={Stull, Roland B},
  year={2012},
  publisher={Springer Science \& Business Media},
}

@article{lettau1969note,
  title={Note on aerodynamic roughness-parameter estimation on the basis of roughness-element description},
  author={Lettau, H\_},
  journal={Journal of Applied Meteorology (1962-1982)},
  volume={8},
  number={5},
  pages={828--832},
  year={1969},
  publisher={JSTOR},
  doi = {10.1175/1520-0450(1969)008<0828:NOARPE>2.0.CO;2}
}

@article{bou2020persistent,
  title={The Persistent Challenge of Surface Heterogeneity in Boundary-Layer Meteorology: A Review},
  author={Bou-Zeid, Elie and Anderson, William and Katul, Gabriel G and Mahrt, Larry},
  journal={Boundary-Layer Meteorology},
  volume={177},
  number={2},
  pages={227--245},
  year={2020},
  publisher={Springer},
  doi = {10.1007/s10546-020-00551-8}
}

@article{wiernga1993representative,
  title={Representative roughness parameters for homogeneous terrain},
  author={Wiernga, Jon},
  journal={Boundary-Layer Meteorology},
  volume={63},
  number={4},
  pages={323--363},
  year={1993},
  publisher={Springer},
  doi = {10.1007/BF00705357}
}

@article{monin1954basic,
  title={Basic laws of turbulent mixing in the surface layer of the atmosphere},
  author={Monin, Andrei Sergeevich and Obukhov, Aleksandr Mikhailovich},
  journal={Contrib. Geophys. Inst. Acad. Sci. USSR},
  volume={151},
  number={163},
  pages={e187},
  year={1954}
}

@article{zilitinkevich2008effect,
  title={The effect of stratification on the aerodynamic roughness length and displacement height},
  author={Zilitinkevich, Sergej S and Mammarella, Ivan and Baklanov, Alexander A and Joffre, Sylvain M},
  journal={Boundary-layer meteorology},
  volume={129},
  number={2},
  pages={179--190},
  year={2008},
  publisher={Springer},
  doi = {10.1007/s10546-008-9307-9}
}

@techreport{krishnamurthy2020boundary,
  title={Boundary layer climatology at arm southern great plains},
  author={Krishnamurthy, Raghavendra and Newsom, Rob K and Chand, Duli and Shaw, William J},
  year={2020},
  institution={Pacific Northwest National Laboratory (PNNL), Richland, WA (United States)}
}

@book{gelman1995bayesian,
  title={Bayesian data analysis},
  author={Gelman, Andrew and Carlin, John B and Stern, Hal S and Rubin, Donald B},
  year={1995},
  publisher={Chapman and Hall/CRC},
  doi = {10.1201/9780429258411}
}

@book{wyngaard2010turbulence,
  title={Turbulence in the Atmosphere},
  author={Wyngaard, John C},
  year={2010},
  publisher={Cambridge university press},
}

@article{heck2025coriolis,
  title={Coriolis effects on wind turbine wakes across neutral atmospheric boundary layer regimes},
  author={Heck, Kirby S and Howland, Michael F},
  journal={Journal of Fluid Mechanics},
  volume={1008},
  pages={A7},
  year={2025},
  publisher={Cambridge University Press},
  doi = {10.1017/jfm.2025.35}
}

@article{bou2005scale,
  title={A scale-dependent Lagrangian dynamic model for large eddy simulation of complex turbulent flows},
  author={Bou-Zeid, Elie and Meneveau, Charles and Parlange, Marc},
  journal={Physics of fluids},
  volume={17},
  number={2},
  year={2005},
  publisher={AIP Publishing},
  doi = {10.1063/1.1839152}
}

@article{ghate2017subfilter,
  title={Subfilter-scale enrichment of planetary boundary layer large eddy simulation using discrete Fourier--Gabor modes},
  author={Ghate, Aditya S and Lele, Sanjiva K},
  journal={Journal of Fluid Mechanics},
  volume={819},
  pages={494--539},
  year={2017},
  publisher={Cambridge University Press},
  doi = {10.1017/jfm.2017.187}
}

@article{nicoud2011using,
  title={Using singular values to build a subgrid-scale model for large eddy simulations},
  author={Nicoud, Franck and Toda, Hubert Baya and Cabrit, Olivier and Bose, Sanjeeb and Lee, Jungil},
  journal={Physics of fluids},
  volume={23},
  number={8},
  year={2011},
  publisher={AIP Publishing},
  doi = {10.1063/1.3623274}
}

@article{shin2025addressing,
  title={Addressing Grid Convergence and Log-Layer Mismatch in Wall Modeled Large Eddy Simulations of Geophysical Flows Over Rough Surfaces and Canopies},
  author={Shin, Ethan Y and Yang, Xiang IA and Howland, Michael F},
  journal={Boundary-Layer Meteorology},
  volume={191},
  number={9},
  pages={42},
  year={2025},
  publisher={Springer},
  doi = {10.1007/s10546-025-00934-9}
}

@article{stoll2020large,
  title={Large-eddy simulation of the atmospheric boundary layer},
  author={Stoll, Rob and Gibbs, Jeremy A and Salesky, Scott T and Anderson, William and Calaf, Marc},
  journal={Boundary-Layer Meteorology},
  volume={177},
  number={2},
  pages={541--581},
  year={2020},
  publisher={Springer},
  doi = {10.1007/s10546-020-00556-3}
}

@article{kumar2006large,
  title={Large-eddy simulation of a diurnal cycle of the atmospheric boundary layer: Atmospheric stability and scaling issues},
  author={Kumar, Vijayant and Kleissl, Jan and Meneveau, Charles and Parlange, Marc B},
  journal={Water resources research},
  volume={42},
  number={6},
  year={2006},
  publisher={Wiley Online Library},
  doi = {10.1029/2005WR004651}
}

@article{moeng1984large,
  title={A large-eddy-simulation model for the study of planetary boundary-layer turbulence},
  author={Moeng, Chin-Hoh},
  journal={Journal of Atmospheric Sciences},
  volume={41},
  number={13},
  pages={2052--2062},
  year={1984},
  doi = {10.1175/1520-0469(1984)041<2052:ALESMF>2.0.CO;2}
}

@article{howland2020influence,
  title={Influence of the geostrophic wind direction on the atmospheric boundary layer flow},
  author={Howland, Michael F and Ghate, Aditya S and Lele, Sanjiva K},
  journal={Journal of Fluid Mechanics},
  volume={883},
  pages={A39},
  year={2020},
  publisher={Cambridge University Press},
  doi = {10.1017/jfm.2019.889}
}

@article{liu2021geostrophic,
  title={Geostrophic drag law for conventionally neutral atmospheric boundary layers revisited},
  author={Liu, Luoqin and Gadde, Srinidhi N and Stevens, Richard JAM},
  journal={Quarterly journal of the royal meteorological society},
  volume={147},
  number={735},
  pages={847--857},
  year={2021},
  publisher={Wiley Online Library},
  doi = {10.1002/qj.3949}
}

@article{foken200650,
  title={50 years of the Monin--Obukhov similarity theory},
  author={Foken, Thomas},
  journal={Boundary-Layer Meteorology},
  volume={119},
  number={3},
  pages={431--447},
  year={2006},
  publisher={Springer},
  doi = {10.1007/s10546-006-9048-6}
}

@article{verkaik2000evaluation,
  title={Evaluation of two gustiness models for exposure correction calculations},
  author={Verkaik, JW},
  journal={Journal of Applied Meteorology},
  volume={39},
  number={9},
  pages={1613--1626},
  year={2000},
  doi = {10.1175/1520-0450(2000)039<1613:EOTGMF>2.0.CO;2}
}

@article{barthelmie1993estimation,
  title={Estimation of sector roughness lengths and the effect on prediction of the vertical wind speed profile},
  author={Barthelmie, RJ and Palutikof, JP and Davies, TD},
  journal={Boundary-Layer Meteorology},
  volume={66},
  number={1},
  pages={19--47},
  year={1993},
  publisher={Springer},
  doi = {10.1007/BF00705458}
}

@article{graf2014intercomparison,
  title={Intercomparison of methods for the simultaneous estimation of zero-plane displacement and aerodynamic roughness length from single-level eddy-covariance data},
  author={Graf, Alexander and van de Boer, Anneke and Moene, Arnold and Vereecken, Harry},
  journal={Boundary-layer meteorology},
  volume={151},
  number={2},
  pages={373--387},
  year={2014},
  publisher={Springer},
  doi = {10.1007/s10546-013-9905-z}
}

@article{gneiting2007strictly,
  title={Strictly proper scoring rules, prediction, and estimation},
  author={Gneiting, Tilmann and Raftery, Adrian E},
  journal={Journal of the American statistical Association},
  volume={102},
  number={477},
  pages={359--378},
  year={2007},
  publisher={Taylor \& Francis},
  doi = {10.1198/016214506000001437}
}

@techreport{osti_3029654,
  author       = {Sullivan, Ryan and Pal, Sujan and Cook, D. R.},
  title        = {Eddy Correlation Flux Measurement System (ECOR) Instrument Handbook},
  institution  = {Oak Ridge National Laboratory (ORNL), Oak Ridge, TN (United States). Atmospheric Radiation Measurement (ARM) User Facility},
  doi          = {10.2172/3029654},
  url          = {https://www.osti.gov/biblio/3029654},
  place        = {United States},
  year         = {2026},
  month        = {04}
}

@article{grimmond1999aerodynamic,
  title={Aerodynamic properties of urban areas derived from analysis of surface form},
  author={Grimmond, C Sue B and Oke, Timothy R},
  journal={Journal of applied meteorology},
  volume={38},
  number={9},
  pages={1262--1292},
  year={1999},
  doi = {10.1175/1520-0450(1999)038<1262:APOUAD>2.0.CO;2}
}

@article{stokes1994atmospheric,
  title={The Atmospheric Radiation Measurement (ARM) Program: Programmatic background and design of the cloud and radiation test bed},
  author={Stokes, Gerald M and Schwartz, Stephen E},
  journal={Bulletin of the American Meteorological Society},
  volume={75},
  number={7},
  pages={1201--1222},
  year={1994},
  publisher={American Meteorological Society},
  doi = {10.1175/1520-0477(1994)075<1201:TARMPP>2.0.CO;2}
}

@article{jackson1981displacement,
  title={On the displacement height in the logarithmic velocity profile},
  author={Jackson, PS},
  journal={Journal of fluid mechanics},
  volume={111},
  pages={15--25},
  year={1981},
  publisher={Cambridge University Press},
  doi = {10.1017/S0022112081002279}
}

@article{anderson2010large,
  title={A large-eddy simulation model for boundary-layer flow over surfaces with horizontally resolved but vertically unresolved roughness elements},
  author={Anderson, William and Meneveau, Charles},
  journal={Boundary-layer meteorology},
  volume={137},
  number={3},
  pages={397--415},
  year={2010},
  publisher={Springer},
  doi = {10.1007/s10546-010-9537-5}
}

@article{sogachev2016displacement,
  title={On displacement height, from classical to practical formulation: stress, turbulent transport and vorticity considerations},
  author={Sogachev, Andrey and Kelly, Mark},
  journal={Boundary-layer meteorology},
  volume={158},
  number={3},
  pages={361--381},
  year={2016},
  publisher={Springer},
  doi = {10.1007/s10546-015-0093-x}
}

@article{mahrt2000surface,
  title={Surface heterogeneity and vertical structure of the boundary layer},
  author={Mahrt, L},
  journal={Boundary-Layer Meteorology},
  volume={96},
  number={1},
  pages={33--62},
  year={2000},
  publisher={Springer},
  doi = {10.1023/A:1002482332477}
}

@article{dyer1974review,
  title={A review of flux-profile relationships},
  author={Dyer, AJꎬ},
  journal={Boundary-Layer Meteorology},
  volume={7},
  number={3},
  pages={363--372},
  year={1974},
  publisher={Springer},
  doi = {10.1007/BF00240838}
}

@book{kaimal1994atmospheric,
  title={Atmospheric boundary layer flows: their structure and measurement},
  author={Kaimal, Jagadish Chandran and Finnigan, John J},
  year={1994},
  publisher={Oxford university press}
}

@article{foken1996tools,
  title={Tools for quality assessment of surface-based flux measurements},
  author={Foken, Th and Wichura, Bodo},
  journal={Agricultural and forest meteorology},
  volume={78},
  number={1-2},
  pages={83--105},
  year={1996},
  publisher={Elsevier},
  doi = {10.1016/0168-1923(95)02248-1}
}

@article{andre1986effective,
  title={On the effective roughness length for use in numerical three-dimensional models},
  author={Andr{\'e}, Jean-Claude and Blondin, Christian},
  journal={Boundary-layer meteorology},
  volume={35},
  number={3},
  pages={231--245},
  year={1986},
  publisher={Springer},
  doi = {10.1007/BF00123642}
}

@article{businger1971flux,
  title={Flux-profile relationships in the atmospheric surface layer},
  author={Businger, Joost A and Wyngaard, John C and Izumi, Yutaka and Bradley, Edward F},
  journal={Journal of Atmospheric Sciences},
  volume={28},
  number={2},
  pages={181--189},
  year={1971},
  doi = {10.1175/1520-0469(1971)028<0181:FPRITA>2.0.CO;2}
}

@article{metzger2008time,
  title={Time scales in the unstable atmospheric surface layer},
  author={Metzger, Meredith and Holmes, Heather},
  journal={Boundary-layer meteorology},
  volume={126},
  number={1},
  pages={29--50},
  year={2008},
  publisher={Springer},
  doi = {10.1007/s10546-007-9219-0}
}

@article{mahrt2020non,
  title={Non-stationary boundary layers},
  author={Mahrt, L and Bou-Zeid, Elie},
  journal={Boundary-Layer Meteorology},
  volume={177},
  number={2},
  pages={189--204},
  year={2020},
  publisher={Springer},
  doi = {10.1007/s10546-020-00533-w}
}

@article{lenschow1994long,
  title={How long is long enough when measuring fluxes and other turbulence statistics?},
  author={Lenschow, DH and Mann, Jakob and Kristensen, Leif},
  journal={Journal of Atmospheric and Oceanic Technology},
  volume={11},
  number={3},
  pages={661--673},
  year={1994},
  publisher={American Meteorological Society},
  doi = {10.1175/1520-0426(1994)011<0661:HLILEW>2.0.CO;2}
}

@article{mason1988formation,
  title={The formation of areally-averaged roughness lengths},
  author={Mason, PJ},
  journal={Quarterly Journal of the Royal Meteorological Society},
  volume={114},
  number={480},
  pages={399--420},
  year={1988},
  publisher={Wiley Online Library},
  doi = {10.1002/qj.49711448007}
}

@article{taylor1969wind,
  title={On wind and shear stress profiles above a change in surface roughness},
  author={Taylor, PA},
  journal={Quarterly Journal of the Royal Meteorological Society},
  volume={95},
  number={403},
  pages={77--91},
  year={1969},
  publisher={Wiley Online Library},
  doi = {10.1002/qj.49709540306}
}

@article{horne2026estimating,
  title={Estimating and Evaluating Roughness Length and Displacement Height in Heterogeneous Urban Environments},
  author={Horne, Jason P and Pan, Ying and Davis, Kenneth J},
  journal={Boundary-Layer Meteorology},
  volume={192},
  number={5},
  pages={25},
  year={2026},
  publisher={Springer},
  doi = {10.1007/s10546-026-00968-7}
}

\end{document}